\documentclass[twocolumn]{jpsj3}
\usepackage{bm}
\usepackage{tabularx}
\usepackage{ascmac}
\usepackage{amsmath}
\usepackage{wrapfig}
\usepackage{graphicx}
\usepackage{float}
\usepackage[FIGBOTCAP]{subfigure}
\usepackage{color}

\bmdefine{\boldb}{b}
\bmdefine{\bolds}{s}
\bmdefine{\boldS}{S}
\bmdefine{\boldso}{so}
\bmdefine{\boldSO}{SO}
\bmdefine{\boldi}{i}
\bmdefine{\boldj}{j}
\bmdefine{\boldtau}{\tau}
\bmdefine{\boldsigma}{\sigma}
\bmdefine{\boldl}{l}
\bmdefine{\boldL}{L}
\bmdefine{\boldnabla}{\nabla}
\bmdefine{\boldlambda}{\lambda}
\bmdefine{\boldx}{x}
\bmdefine{\boldX}{X}
\bmdefine{\boldk}{k}
\bmdefine{\boldK}{K}
\bmdefine{\boldp}{p}
\bmdefine{\boldq}{q}
\bmdefine{\boldQ}{Q}
\bmdefine{\boldD}{D}
\bmdefine{\boldr}{r}
\bmdefine{\boldR}{R}
\bmdefine{\boldn}{n}
\bmdefine{\boldj}{j}
\bmdefine{\boldA}{A}
\bmdefine{\boldzero}{0}
\bmdefine{\boldone}{1}
\bmdefine{\boldtwo}{2}
\bmdefine{\boldthree}{3}
\bmdefine{\boldfour}{4}

\title{Magnon Dispersion and Specific Heat 
of Chiral Magnets on the Pyrochlore Lattice}

\author{Naoya Arakawa}
\inst{Center for Emergent Matter Science (CEMS), 
RIKEN, Wako, Saitama 351-0198, Japan}

\abst{
Chiral magnets are magnetically ordered insulators having spin scalar chirality, 
and 
magnons of chiral magnets have been poorly understood.
We study the magnon dispersion and specific heat 
for four chiral magnets with $\boldQ=\boldzero$ 
on the pyrochlore lattice. 
This study is based on the linear-spin-wave approximation 
for the $S=\frac{1}{2}$ effective Hamiltonian 
consisting of 
two kinds of Heisenberg interaction and 
two kinds of Dzyaloshinsky-Moriya interaction. 
We show that 
the three-in-one-out type chiral magnets 
possess 
an optical branch of the magnon dispersion near $\boldq=\boldzero$, 
in addition to three quasiacoustic branches.
This differs from  
the all-in/all-out type chiral magnets, 
which possess four quasiacoustic branches.  
We also show that 
all four chiral magnets have 
a gapped magnon energy at $\boldq=\boldzero$, 
indicating the absence of the Goldstone type gapless excitation. 
These results are useful for experimentally identifying 
the three-in-one-out or all-in/all-out type chiral order.
Then, 
we show that 
there is no qualitative difference in the specific heat 
among the four magnets. 
This indicates that 
the specific heat is not useful for distinguishing the kinds of chiral orders. 
We finally compare our results with experiments 
and provide a proposal for the three-in-one-out type chiral magnets. 
}

\begin{document}
\maketitle

\section{Introduction}

Magnons are bosonic quasiparticles 
describing the low-energy properties of a magnetically ordered system~\cite{Bloch,Slater}. 
A magnetically ordered system 
has a finite expectation value of some component of the spin 
at each site. 
The effects of the spin fluctuations can be described 
in terms of the bosonic operators 
because the spin operators can be expressed in terms of those
using the Holstein-Primakoff transformation~\cite{HP,Dyson1,Dyson2}. 
Since these bosonic operators describe the creation and annihilation of a magnon 
and the spin Hamiltonian is expressed in terms of these magnon operators, 
magnons describe the low-energy properties of a magnet. 
Note that 
the zeroth-order terms of the magnon operators describe the zero-point motion, 
the second-order terms describe the noninteracting magnons, 
and the higher-order terms describe the interactions between magnons. 

Comparing with the understanding of magnons of nonchiral magnets, 
the low-temperature properties of magnons of chiral magnets are poorly understood. 
A chiral magnet is a magnetically ordered insulator 
with spin scalar chirality, 
while a nonchiral magnet does not have spin scalar chirality. 
(Our definition is distinct from another definition, 
in which a magnet on a chiral lattice is a chiral magnet, 
because our definition is applicable to even a nonchiral lattice, 
such as the pyrochlore lattice.)
It is widely known~\cite{Yosida-text,Kanamori-text} that 
a major difference between collinear ferromagnetic (FM) and antiferromagnetic (AF) 
magnets, typical nonchiral magnets, 
is the different curvature of the magnon dispersion 
in the long-wavelength limit, 
and  
this curvature difference 
causes a difference in the temperature dependence of 
some quantity, such as the specific heat. 
However, 
such knowledge about chiral magnets is unsatisfactory 
even for noninteracting magnons. 
Thus, 
a detailed theoretical analysis of the chiral magnets may be needed.

\begin{figure}[tb]
\includegraphics[width=76mm]{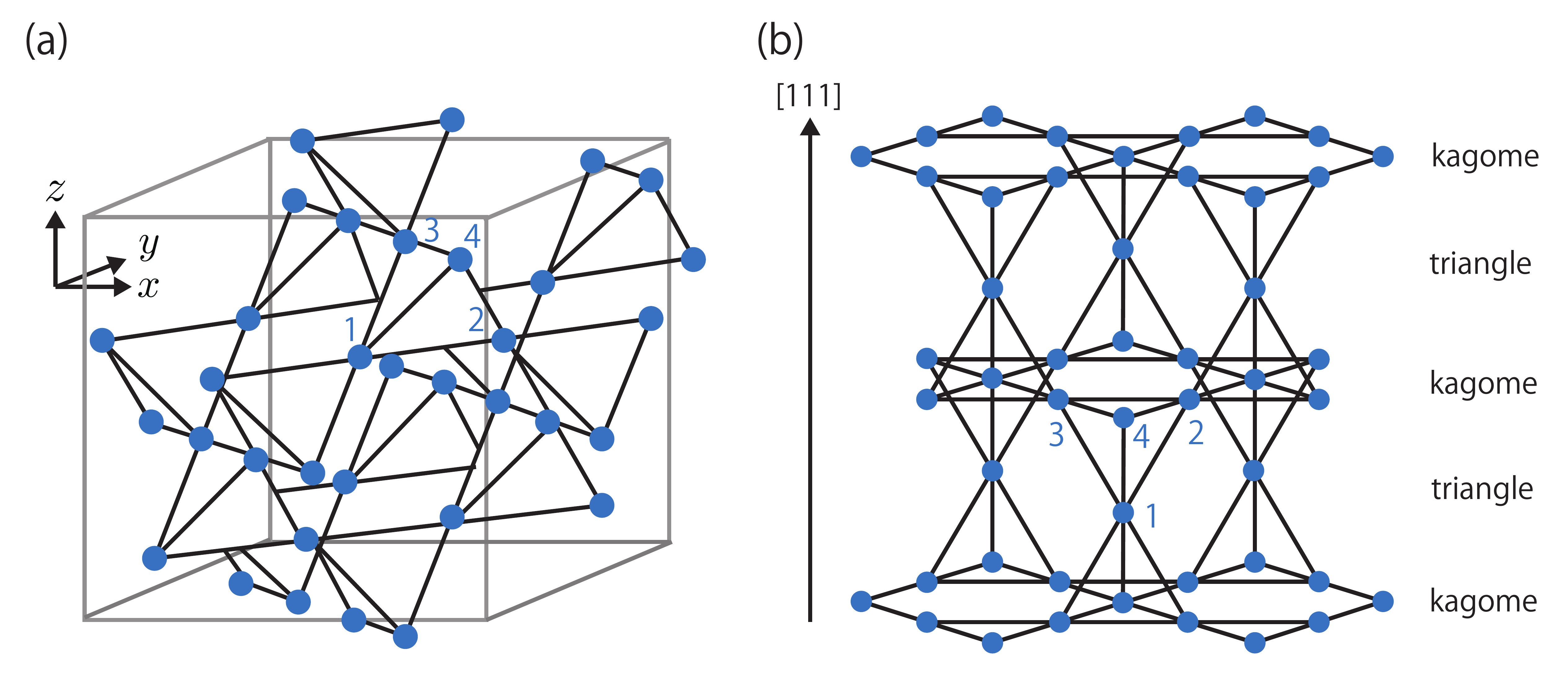}
\vspace{-10pt}
\caption{
(Color online)
(a) Pyrochlore lattice consisting of B ions in A$_{2}$B$_{2}$O$_{7}$ or AB$_{2}$O$_{4}$, 
and (b) the lattice as alternating kagome and triangle layers 
along the $[111]$ direction. 
Blue circles represent B ions, and 
$1$, $2$, $3$, and $4$ denote sublattices 1, 2, 3, and 4, 
respectively. 
}
\label{fig1}
\end{figure}

\begin{figure}[tb]
\includegraphics[width=64mm]{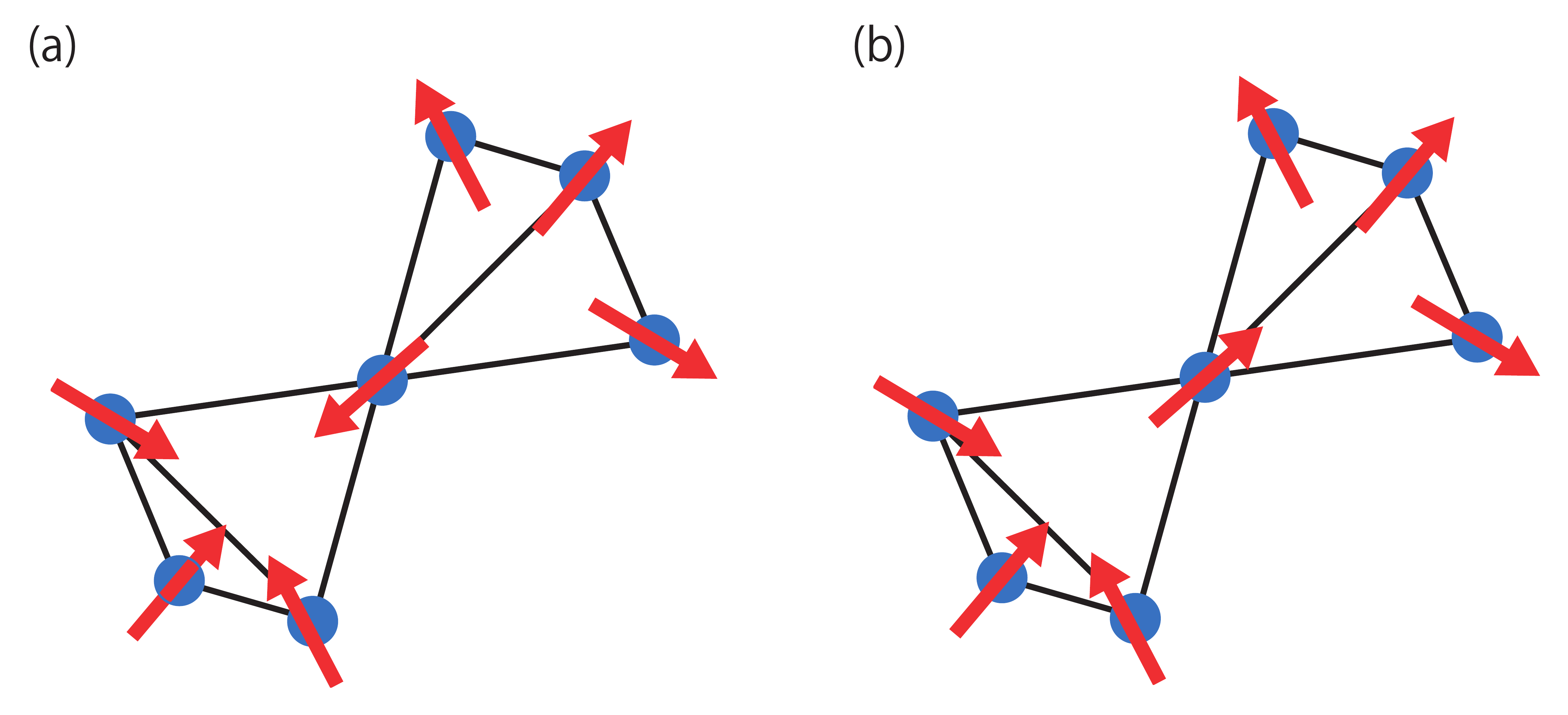}
\vspace{-10pt}
\caption{
(Color online) Spin structures of 
(a) the AIAO and (b) 3I1O chiral magnets. 
Their ordering vectors are $\boldQ=\boldzero$. 
}
\label{fig2}
\end{figure}

To understand the properties of magnons of chiral magnets, 
we focus on $S=\frac{1}{2}$ pyrochlore magnets,  
magnetically ordered insulators on the pyrochlore lattice~\cite{Pyro-review} (Fig. \ref{fig1})
with $S=\frac{1}{2}$ per site.  
The pyrochlore magnets can be chiral, 
depending on the values of exchange interactions. 
Examples of chiral magnets are 
all-in/all-out (AIAO)~\cite{AIAO-exp} and 
three-in-one-out (3I1O) ones~\cite{3I1O-theory}; 
their spin structures are shown in Fig. \ref{fig2}. 
Although the magnon properties of the AIAO chiral magnet have been revealed~\cite{AIAO-exp}, 
it is unclear 
how they differ from magnon properties of other chiral magnets, 
and what property is characteristic of chiral magnets. 
It is thus desirable to clarify differences and similarities 
between magnons of the AIAO type and 3I1O type chiral magnets.

In this paper, 
we study the low-temperature properties 
of noninteracting magnons 
of $S=\frac{1}{2}$ pyrochlore magnets 
using the linear-spin-wave approximation (LSWA)~\cite{LSWA1,LSWA2}. 
We consider 
several chiral magnets with $\boldQ=\boldzero$, 
where $\boldQ$ is the ordering vector. 
We show that 
the 3I1O type chiral magnets 
possess 
an optical branch of the magnon dispersion 
near $\boldq=\boldzero$, 
in addition to three quasiacoustic branches, 
while 
the four branches in the AIAO type chiral magnets 
are all quasiacoustic. 
We also show that 
the magnon energy at $\boldq=\boldzero$ has a finite gap 
for the AIAO type and the 3I1O type chiral magnets. 

Then, 
we show the results for the ground state of $S=\frac{1}{2}$ pyrochlore magnets 
using the mean-field approximation (MFA) for several chiral or nonchiral orders 
with $\boldQ=\boldzero$. 
This is because the ground state determined in the MFA 
is a starting point to include the effects of fluctuations. 

We organize this paper as follows. 
In Sect. 2, 
we explain our method for chiral magnets with $\boldQ=\boldzero$ 
in pyrochlore oxides with weak spin-orbit coupling (SOC). 
In Sect. 3, 
we show the results obtained using the MFA 
and the LSWA. 
In Sect. 4, 
we discuss the applicability of the LSWA 
and the correspondences between our results and several experimental results, 
and propose several observable aspects for future experiments. 
In Sect. 5, 
we summarize our main results. 

\section{Method}
In this section, 
we show our low-energy effective model for the $S=\frac{1}{2}$ pyrochlore oxides 
with weak SOC, 
and explain the MFA and the LSWA. 
In Sect. 2.1, 
we introduce the spin Hamiltonian consisting 
of the Heisenberg and Dzyaloshinsky-Moriya (DM) 
interactions between 
nearest-neighbor (NN) B ions on the pyrochlore lattice. 
In Sect. 2.2, 
we formulate the MFA for our spin Hamiltonian 
and explain how to determine the ground state. 
In Sect. 2.3, 
we formulate the LSWA 
for a commensurate magnetic order 
and then show an algorithm to obtain numerical solutions of the LSWA. 
Throughout this paper, 
we set $\hbar=1$ and $k_{\textrm{B}}=1$.

\subsection{Model}

As an effective Hamiltonian for the $S=\frac{1}{2}$ pyrochlore magnets, 
we use the following spin Hamiltonian~\cite{NA-pyro}: 
\begin{align}
\hat{H}_{\textrm{eff}}
=&\sum\limits_{\langle \boldi,\boldj \rangle}
J_{\boldi \boldj}\hat{\boldS}_{\boldi}\cdot \hat{\boldS}_{\boldj}
+\sum\limits_{\langle \boldi,\boldj \rangle}
\boldD_{\boldi \boldj}\cdot (\hat{\boldS}_{\boldi}\times \hat{\boldS}_{\boldj})\notag\\
=&
\sum\limits_{m,n=1}^{N}
\sum\limits_{l,l^{\prime}=1}^{4}
\sum\limits_{\alpha,\beta=x,y,z}
M_{mlnl^{\prime}}^{\alpha\beta}
\hat{S}_{\boldR_{m}+\boldr_{l}}^{\alpha} 
\hat{S}_{\boldR_{n}+\boldr_{l^{\prime}}}^{\beta}
.\label{eq:Heff}
\end{align}
Here  
$\boldi$ and $\boldj$ are site indices, 
$\boldR_{m}$ and $\boldR_{n}$ are translational vectors, 
$\boldr_{l}$ and $\boldr_{l^{\prime}}$ are sublattice vectors 
[i.e., $\boldr_{1}={}^{t}(0\ 0\ 0)$,
$\boldr_{2}={}^{t}(\frac{1}{2}\ \frac{1}{2}\ 0)$, 
$\boldr_{3}={}^{t}(0\ \frac{1}{2}\ \frac{1}{2})$, 
and $\boldr_{4}={}^{t}(\frac{1}{2}\ 0\ \frac{1}{2})$], and 
$J_{\boldi \boldj}$ and $\boldD_{\boldi \boldj}$ are 
the Heisenberg and DM interactions between the NN B sites, respectively. 
$J_{\boldi \boldj}=J_{\boldj \boldi}$ and 
$\boldD_{\boldi \boldj}=-\boldD_{\boldj \boldi}$ are given by
\begin{align}
&J_{\boldone \boldtwo}
= J_{0}+J_{1},  
\boldD_{\boldone \boldtwo}
=\left(
\begin{array}{@{\,}c@{\,}}
-D_{0}-D_{1}\\
+D_{0}+D_{1}\\
0
\end{array} 
\right),\label{eq:J12}\\
&J_{\boldone \boldthree}
= J_{0}+J_{1},
\boldD_{\boldone \boldthree}
=
\left(
\begin{array}{@{\,}c@{\,}}
0\\
-D_{0}-D_{1}\\
+D_{0}+D_{1}
\end{array} 
\right),\label{eq:J13}\\
&J_{\boldone \boldfour}
= J_{0}+J_{1}, 
\boldD_{\boldone \boldfour}
=
\left(
\begin{array}{@{\,}c@{\,}}
+D_{0}+D_{1}\\
0\\
-D_{0}-D_{1}
\end{array} 
\right),\label{eq:J14}\\
&J_{\boldfour \boldthree}
= J_{0}-J_{1},
\boldD_{\boldfour \boldthree}
=
\left(
\begin{array}{@{\,}c@{\,}}
+D_{0}-D_{1}\\
+D_{0}-D_{1}\\
0
\end{array} 
\right),\label{eq:J43}\\
&J_{\boldtwo \boldfour}
= J_{0}-J_{1},
\boldD_{\boldtwo \boldfour}
=
\left(
\begin{array}{@{\,}c@{\,}}
0\\
+D_{0}-D_{1}\\
+D_{0}-D_{1}
\end{array} 
\right),\label{eq:J24}\\
&J_{\boldtwo \boldthree}
= J_{0}-J_{1}, 
\boldD_{\boldtwo \boldthree}
=\left(
\begin{array}{@{\,}c@{\,}}
-D_{0}+D_{1}\\
0\\
-D_{0}+D_{1}
\end{array} 
\right).\label{eq:J23}
\end{align}
The above $\boldD_{\boldi \boldj}$ are consistent with the Moriya rule~\cite{DM-Moriya,DMI-pyro}. 

In our effective model, 
we consider not only the $J_{0}$ and $D_{0}$ terms 
but also the $J_{1}$ and $D_{1}$ terms for the following four reasons. 
First, 
the model only with the $J_{1}$ and $D_{1}$ terms is a minimal model 
for the 3I1O chiral magnet. 
The 3I1O chiral magnet is the most stable ground state in the minimal model 
for the negative $J_{1}$ and $D_{1}$ (see Table II). 
Second, 
the magnon properties obtained in the minimal model 
may remain qualitatively unchanged in a more realistic model for the 3I1O chiral magnet. 
Third, 
the model with the $J_{0}$, $D_{0}$, $J_{1}$, and $D_{1}$ terms 
is useful for analyzing the magnon properties of another chiral magnet 
whose spin structure is similar to that of the AIAO or 3I1O chiral magnet. 
A chiral magnet similar to the AIAO one is a distorted AIAO (dAIAO) chiral magnet, 
and a chiral magnet similar to the 3I1O one is a distorted 3I1O (d3I1O) chiral magnet 
(see Table III). 
Such analyses are useful for understanding magnon properties 
of the AIAO type and 3I1O type chiral magnets. 
Fourth,  
the values of the $J_{1}$ and $D_{1}$ terms may be controllable, 
for example, by varying the pressure along the $[111]$ direction. 

\subsection{Mean-field approximation}

We start with the formulation of the MFA. 
In the MFA, 
we neglect fluctuations and replace one of the two spin operators in the Hamiltonian 
by the expectation value. 
Thus, 
the ground state in the MFA is given by 
\begin{align}
\langle\hat{H}_{\textrm{eff}}\rangle
=&
\sum\limits_{m,n=1}^{N}
\sum\limits_{l,l^{\prime}=1}^{4}
\sum\limits_{\alpha,\beta}
M_{mlnl^{\prime}}^{\alpha\beta}
\langle\hat{S}_{\boldR_{m}+\boldr_{l}}^{\alpha}\rangle 
\langle\hat{S}_{\boldR_{n}+\boldr_{l^{\prime}}}^{\beta}\rangle.\label{eq:Heff-MFA-r}
\end{align}
In the above equation, 
$\langle\hat{\boldS}_{\boldi}\rangle$ should satisfy 
the hard-spin constraint, i.e., $S=\frac{1}{2}=|\langle\hat{\boldS}_{\boldi}\rangle|$. 
Since Eq. (\ref{eq:Heff-MFA-r}) can be expressed as a hermitian form 
in terms of the Fourier coefficients of  
$\langle\hat{S}^{\alpha}_{\boldi}\rangle$ and 
$\langle\hat{S}^{\beta}_{\boldj}\rangle$, 
\begin{align}
\langle \hat{H}_{\textrm{eff}}\rangle
=
\sum\limits_{\boldq}
\sum\limits_{l,l^{\prime}=1}^{4}
\sum\limits_{\alpha,\beta}
M_{ll^{\prime}}^{\alpha \beta}(\boldq)
\langle \hat{S}_{\boldq l^{\prime}}^{\beta}\rangle 
\langle \hat{S}_{\boldq l}^{\alpha}\rangle^{\ast},\label{eq:HMFA}
\end{align}
the determination of the ground state is equivalent to 
the determination of the minimum of 
$\langle \hat{H}_{\textrm{eff}}\rangle$ as a function of $\boldq$ 
under the hard-spin constraint. 
In Eq. (\ref{eq:HMFA}), 
we have introduced 
$M_{ll^{\prime}}^{\alpha \beta}(\boldq)=M_{l^{\prime}l}^{\beta\alpha}(\boldq)$ [$M_{ll}^{\alpha \beta}(\boldq)=0$], 
given by 
\begin{align}
&M_{21}^{\alpha \beta}(\boldq)
=Z_{+}^{\alpha\beta}\cos(\frac{q_{x}}{2}+\frac{q_{y}}{2}),\\
&M_{43}^{\alpha \beta}(\boldq)
=Z_{-}^{\alpha\beta}\cos(\frac{q_{x}}{2}-\frac{q_{y}}{2}),\\
&M_{31}^{\alpha \beta}(\boldq)
=X_{+}^{\alpha\beta}\cos(\frac{q_{y}}{2}+\frac{q_{z}}{2}),\\
&M_{42}^{\alpha \beta}(\boldq)
=X_{-}^{\alpha\beta}\cos(\frac{q_{y}}{2}-\frac{q_{z}}{2}),\\
&M_{41}^{\alpha \beta}(\boldq)
=Y_{+}^{\alpha\beta}\cos(\frac{q_{z}}{2}+\frac{q_{x}}{2}),\\
&M_{32}^{\alpha \beta}(\boldq)
=Y_{-}^{\alpha\beta}\cos(\frac{q_{z}}{2}-\frac{q_{x}}{2}),
\end{align}
with
\begin{align} 
&Z_{\pm}^{\alpha\beta}
=
\begin{cases}
J_{0}\pm J_{1} & \textrm{for}\ \alpha=\beta \\
D_{0}\pm D_{1} & \textrm{for}\ (\alpha,\beta)=(y,z)\\
\pm D_{0}+ D_{1} & \textrm{for}\ (\alpha,\beta)=(x,z)\\
-(D_{0}\pm D_{1}) & \textrm{for}\ (\alpha,\beta)=(z,y)\\
-(\pm D_{0}+ D_{1}) & \textrm{for}\ (\alpha,\beta)=(z,x)
\end{cases},\\
&X_{\pm}^{\alpha\beta}
=
\begin{cases}
J_{0}\pm J_{1} & \textrm{for}\ \alpha=\beta \\
D_{0}\pm D_{1} & \textrm{for}\ (\alpha,\beta)=(y,x)\\
\pm D_{0}+ D_{1} & \textrm{for}\ (\alpha,\beta)=(z,x)\\
-(D_{0}\pm D_{1}) & \textrm{for}\ (\alpha,\beta)=(x,y)\\
-(\pm D_{0}+ D_{1}) & \textrm{for}\ (\alpha,\beta)=(x,z)
\end{cases},\\
&Y_{\pm}^{\alpha\beta}
=
\begin{cases}
J_{0}\pm J_{1} & \textrm{for}\ \alpha=\beta \\
D_{0}\pm D_{1} & \textrm{for}\ (\alpha,\beta)=(x,y)\\
\pm D_{0}+ D_{1} & \textrm{for}\ (\alpha,\beta)=(z,y)\\
-(D_{0}\pm D_{1}) & \textrm{for}\ (\alpha,\beta)=(y,x)\\
-(\pm D_{0}+D_{1}) & \textrm{for}\ (\alpha,\beta)=(y,z)
\end{cases}.
\end{align}

\subsection{Linear-spin-wave approximation}

We can formulate the LSWA 
for a commensurate magnetic order in our effective spin Hamiltonian 
in a similar way to Ref. 11. 
Since the spin directions at sites in a unit cell 
can differ even for a commensurate magnetic order, 
we introduce the following rotation, which enables the spin directions to be the same: 
\begin{align}
\hat{S}_{\boldR_{m}+\boldr_{l}}^{\alpha}=
\sum\limits_{\beta=x,y,z}
(R_{l})^{\alpha\beta}\hat{S}_{\boldR_{m}+\boldr_{l}}^{\prime \beta},\label{eq:rot-S}
\end{align}
where $\hat{\boldS}_{\boldi}^{\prime}={}^{t}(0\ 0\ S)$
and 
$(R_{l})^{\alpha\beta}$ for $l=1,2,3,4$ is the rotation matrix. 
$(R_{l})^{\alpha\beta}$ is given by 
the Rodrigues formula for a rotation matrix: 
$(R_{l})^{\alpha\alpha}=\cos \phi_{l} +(n_{l}^{\alpha})^{2}(1-\cos \phi_{l})$, 
$(R_{l})^{xy}=(R_{l})^{yx}=n_{l}^{x}n_{l}^{y}(1-\cos \phi_{l})$, 
$(R_{l})^{xz}=-(R_{l})^{zx}=n_{l}^{y}\sin\phi_{l}$, 
and 
$(R_{l})^{yz}=-(R_{l})^{zy}=-n_{l}^{x}\sin\phi_{l}$
with
$\cos\phi_{l}
=
\frac{\langle \hat{S}_{\boldi}^{z}\rangle}{S}$, 
$\sin\phi_{l}
=
\frac{\sqrt{\langle \hat{S}_{\boldi}^{x}\rangle^{2}+\langle \hat{S}_{\boldi}^{y}\rangle^{2}}}{S}$, and 
$n_{l}^{x}=-
\frac{1}{\sqrt{\langle \hat{S}_{\boldi}^{x}\rangle^{2}+\langle \hat{S}_{\boldi}^{y}\rangle^{2}}}
\langle \hat{S}_{\boldi}^{y}\rangle$, 
$n_{l}^{y}=
\frac{1}{\sqrt{\langle \hat{S}_{\boldi}^{x}\rangle^{2}+\langle \hat{S}_{\boldi}^{y}\rangle^{2}}}
\langle \hat{S}_{\boldi}^{x}\rangle$, 
and 
$n_{l}^{z}=0$.
For commensurate magnetic orders, 
$\langle \hat{\boldS}_{\boldi}\rangle$ 
is independent of $\boldR_{m}$ 
and depends on $\boldr_{l}$. 
By substituting Eq. (\ref{eq:rot-S}) into Eq. (\ref{eq:Heff}), 
we express our effective spin Hamiltonian as 
\begin{align}
\hat{H}_{\textrm{eff}}
=
\sum\limits_{m,n=1}^{N}
\sum\limits_{l,l^{\prime}=1}^{4}
\sum\limits_{\alpha,\beta=x,y,z}
M_{mlnl^{\prime}}^{\prime \alpha\beta}
\hat{S}_{\boldR_{m}+\boldr_{l}}^{\prime\alpha}
\hat{S}_{\boldR_{n}+\boldr_{l^{\prime}}}^{\prime\beta},\label{eq:Heff-rewrote}
\end{align}
with
$M_{mlnl^{\prime}}^{\prime \alpha\beta}
=\sum\limits_{\alpha^{\prime},\beta^{\prime}=x,y,z}
(R_{l})^{\alpha^{\prime}\alpha}M_{mlnl^{\prime}}^{\alpha^{\prime}\beta^{\prime}}
(R_{l})^{\beta^{\prime}\beta}$.
Since $\boldS^{\prime}_{\boldi}$ for all $l$ has one FM component, 
we can apply the Holstein-Primakoff method for a FM order 
to the case of a commensurate order for Eq. (\ref{eq:Heff-rewrote}).
By carrying out this procedure, 
we obtain the following quadratic spin Hamiltonian in terms of 
magnon operators (see Appendix):
\begin{align}
\hat{H}_{\textrm{LSW}}
=&
\sum\limits_{\boldq}
\hat{\boldx}^{\dagger}_{\boldq}H(\boldq)\hat{\boldx}_{\boldq}\notag\\
=&
S\sum\limits_{\boldq}
(\hat{\boldb}_{\boldq}^{\dagger}\ \hat{\boldb}_{-\boldq})
\left(
\begin{array}{@{\,}cc@{\,}}
A(\boldq) & B(\boldq)\\
B^{\ast}(-\boldq) & A^{\ast}(-\boldq)
\end{array} 
\right)
\left(
\begin{array}{@{\,}c@{\,}}
\hat{\boldb}_{\boldq}\\
\hat{\boldb}_{-\boldq}^{\dagger}
\end{array} 
\right).\label{eq:HLSW}
\end{align}
Here 
$\hat{\boldb}_{\boldq}={}^{t}(\hat{b}_{\boldq 1}\ \hat{b}_{\boldq 2}\ \hat{b}_{\boldq 3}\ \hat{b}_{\boldq 4})$ 
and $\hat{\boldb}_{-\boldq}^{\dagger}={}^{t}(\hat{b}_{-\boldq 1}^{\dagger}\ \hat{b}_{-\boldq 2}^{\dagger}\ 
\hat{b}_{-\boldq 3}^{\dagger}\ \hat{b}_{-\boldq 4}^{\dagger})$ 
are the vectors of the annihilation and creation operators of a magnon, respectively; 
$A(\boldq)=[A_{ll^{\prime}}(\boldq)]$ 
and $B(\boldq)=[B_{ll^{\prime}}(\boldq)]$
are the $4\times 4$ matrices given by 
\begin{align}
&A_{ll^{\prime}}(\boldq)
=
\frac{1}{2}[M_{ll^{\prime}}^{\prime xx}(-\boldq)+M_{ll^{\prime}}^{\prime yy}(-\boldq)]\notag\\
&-\frac{i}{2}[M_{ll^{\prime}}^{\prime xy}(-\boldq)-M_{ll^{\prime}}^{\prime yx}(-\boldq)]
-\delta_{l,l^{\prime}}
\sum\limits_{l^{\prime\prime}=1}^{4}M_{ll^{\prime\prime}}^{\prime zz}(\boldzero),
\end{align}
and
\begin{align}
&B_{ll^{\prime}}(\boldq)
=
\frac{1}{2}[M_{ll^{\prime}}^{\prime xx}(-\boldq)-M_{ll^{\prime}}^{\prime yy}(-\boldq)]\notag\\
&+\frac{i}{2}[M_{ll^{\prime}}^{\prime xy}(-\boldq)+M_{ll^{\prime}}^{\prime yx}(-\boldq)],
\end{align} 
respectively, 
with $M_{ll^{\prime}}^{\prime \alpha\beta}(\boldq)$ 
the Fourier coefficient of $M_{mlnl^{\prime}}^{\prime \alpha\beta}$. 
This quadratic spin Hamiltonian, 
Eq. (\ref{eq:HLSW}), describes the low-energy properties of magnons 
in the LSWA. 
Owing to the commutation relations for $\hat{b}_{\boldq l}$ and $\hat{b}^{\dagger}_{\boldq l}$, 
$\hat{\boldx}^{\dagger}_{\boldq}$ and $\hat{\boldx}_{\boldq}$ should satisfy 
\begin{align}
[\hat{\boldx}_{\boldq},\hat{\boldx}_{\boldq}^{\dagger}]
\equiv \hat{\boldx}_{\boldq}{}^{t}(\hat{\boldx}^{\ast}_{\boldq})
-{}^{t}(\hat{\boldx}^{\ast}_{\boldq}{}^{t}\hat{\boldx}_{\boldq})
=g,\label{eq:x-comm}
\end{align}
where $g=(g_{LL^{\prime}})$ is the $8\times 8$ paraunit matrix, 
$g_{LL^{\prime}}=\delta_{L,L^{\prime}}$ for $1\le L,L^{\prime}\le 4$, 
$g_{LL^{\prime}}=-\delta_{L,L^{\prime}}$ for $5\le L,L^{\prime}\le 8$.

To analyze the low-energy properties of magnons in the LSWA, 
we need to diagonalize $H(\boldq)$ in Eq. (\ref{eq:HLSW}) 
and find the eigenvalue 
and the eigenfunction; 
the former and latter respectively give the dispersion 
and wave function of noninteracting magnons. 
(In contrast to the diagonalization for the MFA, 
we need to treat the bosonic operators in the diagonalization for the LSWA; 
as a result, we need to use the paraunitary matrix.)  
After the diagonalization, 
Eq. (\ref{eq:HLSW}) becomes 
\begin{align}
\hat{H}_{\textrm{LSW}}
=&
\sum\limits_{\boldq}
\hat{\boldx}^{\prime\dagger}_{\boldq}E(\boldq)\hat{\boldx}_{\boldq}^{\prime},
\end{align}
where $\hat{\boldx}_{\boldq}^{\prime\dagger}
=(\hat{\boldb}_{\boldq}^{\prime \dagger}\ \hat{\boldb}_{-\boldq}^{\prime})$  
and 
$\hat{\boldx}_{\boldq}^{\prime}={}^{t}(\hat{\boldb}_{\boldq}^{\prime}
\hat{\boldb}_{-\boldq}^{\prime\dagger})$ 
are given by 
$\hat{\boldx}_{\boldq}^{\dagger}=\hat{\boldx}^{\prime \dagger}_{\boldq}P_{\boldq}^{\dagger}$ 
and 
$\hat{\boldx}_{\boldq}=P_{\boldq}\hat{\boldx}^{\prime}_{\boldq}$
with the $8\times 8$ matrix $P_{\boldq}=(P_{L \nu;\boldq})$,  
and $E(\boldq)=[\delta_{\nu,\nu^{\prime}}\epsilon_{\nu}(\boldq)]$ is given by 
\begin{align}
P_{\boldq}^{\dagger}H(\boldq)P_{\boldq}=E(\boldq).\label{eq:mag-dis}
\end{align}
Since $\hat{\boldx}_{\boldq}^{\prime}$ and $\hat{\boldx}_{\boldq}^{\prime\dagger}$ are bosonic operators, 
these should satisfy the commutation relation 
\begin{align}
[\hat{\boldx}^{\prime}_{\boldq},\hat{\boldx}_{\boldq}^{\prime \dagger}]
\equiv \hat{\boldx}^{\prime}_{\boldq}{}^{t}(\hat{\boldx}^{\prime \ast}_{\boldq})
-{}^{t}(\hat{\boldx}^{\prime \ast}_{\boldq}{}^{t}\hat{\boldx}^{\prime}_{\boldq})
=g.\label{eq:x'-comm}
\end{align}
As a result of Eqs. (\ref{eq:x-comm}) and (\ref{eq:x'-comm}), 
$P_{\boldq}$ should satisfy 
\begin{align}
P_{\boldq}gP^{\dagger}_{\boldq}=g.\label{eq:def-para}
\end{align}
This equation is regarded as the definition of a paraunitary matrix and  
differs from the definition of a unitary matrix, 
$U_{\boldq}U_{\boldq}^{\dagger}=U_{\boldq}1U_{\boldq}^{\dagger}=1$ 
with $1$ a unit matrix. 

The diagonalization of $H(\boldq)$ 
using the paraunitary matrix can be carried out 
by the procedure proposed by Colpa\cite{Colpa}. 
Assuming that 
$H(\boldq)$ is positive definite, 
we apply 
the Cholesky decomposition to the matrix $H(\boldq)$: 
$H(\boldq)=K^{\dagger}_{\boldq}K_{\boldq}$,
where $K^{\dagger}_{\boldq}$ and $K_{\boldq}$ are 
the upper and lower triangle $8\times 8$ matrices, respectively. 
If $H(\boldq)$ is semipositive definite, 
we add a tiny positive convergence factor $\Delta$ 
to the diagonal components of $H(\boldq)$ 
to enable $H_{L L^{\prime}}(\boldq)+\Delta \delta_{L,L^{\prime}}$ 
to be positive-definite. 
As a result of the Cholesky decomposition, 
Eq. (\ref{eq:mag-dis}) becomes
\begin{align}
P_{\boldq}^{\dagger}
K^{\dagger}_{\boldq}K_{\boldq}
P_{\boldq}=E(\boldq).
\end{align}
This symmetric form makes the method of finding $E(\boldq)$ and $P_{\boldq}$ easier 
because both are obtained by the diagonalization 
of the $8\times 8$ matrix $K_{\boldq}gK^{\dagger}_{\boldq}$ 
using the unitary matrix in the following way. 
We can diagonalize $K_{\boldq}gK^{\dagger}_{\boldq}$ as follows
by using the $8\times 8$ unitary matrix $U_{\boldq}$: 
\begin{align}
U^{\dagger}_{\boldq}
(K_{\boldq}gK^{\dagger}_{\boldq})
U_{\boldq}=L(\boldq).
\end{align}
The $8\times 8$ diagonal matrix $L(\boldq)$ is connected with $E(\boldq)$, 
and $U_{\boldq}$ is connected with $P_{\boldq}$:
$L(\boldq)=gE(\boldq)$ and $U_{\boldq}=K_{\boldq}P_{\boldq}E(\boldq)^{-\frac{1}{2}}$,
where the $8\times 8$ diagonal matrix $E(\boldq)^{-\frac{1}{2}}$ 
is given by $E_{\nu \nu^{\prime}}(\boldq)^{-\frac{1}{2}}
=\delta_{\nu,\nu^{\prime}}\epsilon_{\nu}(\boldq)^{-\frac{1}{2}}$. 
Since the numerical diagonalization using the unitary matrix is easier 
than that using the paraunitary matrix, 
the above procedure provides a more useful algorithm. 

In the actual calculations for the LSWA, shown in Sect. 3.2, 
we use a tiny positive $\Delta$. 
This is for three reasons: 
one is that 
it is difficult to check whether or not 
the matrix $H(\boldq)$ is positive definite 
before the diagonalization; 
another is that 
we can check whether the matrix $H(\boldq)$ is positive definite  
by checking the obtained eigenvalues, 
which should all be positive for the positive definite case; 
the other is that 
the effects of $\Delta$ on the results shown in this paper 
are negligible at the scale of the exchange interactions. 

\section{Results}
\begin{table*}[tb]
\vspace{-15pt}
\caption{Ground-state 
spin configurations and energies per tetrahedron 
in the MFA for four nonchiral orders.
$S_{1}$ and $S_{2}$ are given by 
$(S_{1})^{2}=S^{2}=\frac{1}{4}$ and $2(S_{2})^{2}=S^{2}=\frac{1}{4}$, 
respectively. }
\label{tab:nonchiral}
\vspace{5pt}
\begin{tabular}{ccccccccc} \hline\\[-10pt]
\multicolumn{1}{c}{ \ \ \ \ \ \ \ \ \ \ } &
\multicolumn{1}{c}{ \ \ Collinear FM  \ \ } & 
\multicolumn{1}{c}{ \ \ Collinear AF \ \ } &
\multicolumn{1}{c}{ \ \ Coplanar AF1 \ \ }&
\multicolumn{1}{c}{ \ \ Coplanar AF2 \ \ } \\[1pt] \hline\\[-5pt]
$\langle \hat{\boldS}_{\boldr_{1}}\rangle$ 
&${}^{t}(0\ 0\ S_{1})$ 
&${}^{t}(0\ 0\ S_{1})$
&${}^{t}(S_{2}\ -S_{2}\ 0)$
&${}^{t}(S_{2}\ -S_{2}\ 0)$ \\[5pt]
$\langle \hat{\boldS}_{\boldr_{2}}\rangle$ 
&${}^{t}(0\ 0\ S_{1})$ 
&${}^{t}(0\ 0\ S_{1})$
&${}^{t}(-S_{2}\ S_{2}\ 0)$
&${}^{t}(-S_{2}\ S_{2}\ 0)$ \\[5pt]
$\langle \hat{\boldS}_{\boldr_{3}}\rangle$ 
&${}^{t}(0\ 0\ S_{1})$ 
&${}^{t}(0\ 0\ -S_{1})$
&${}^{t}(-S_{2}\ -S_{2}\ 0)$
&${}^{t}(S_{2}\ S_{2}\ 0)$ \\[5pt]
$\langle \hat{\boldS}_{\boldr_{4}}\rangle$ 
&${}^{t}(0\ 0\ S_{1})$ 
&${}^{t}(0\ 0\ -S_{1})$
&${}^{t}(S_{2}\ S_{2}\ 0)$
&${}^{t}(-S_{2}\ -S_{2}\ 0)$ \\[5pt]\hline \\[-10pt] 
$\frac{\langle \hat{H}_{\textrm{eff}}\rangle}{N}$ 
&$3J_{0}$ 
&$-J_{0}$
&$-J_{0}-2D_{0}$
&$-J_{0}+2D_{0}$ \\[5pt]\hline 
\end{tabular}
\vspace{-10pt}
\end{table*}
In this section, 
we show the results for the $S=\frac{1}{2}$ pyrochlore magnets. 
In Sect. 3.1, 
we show the ground-state spin configurations, 
energies, and spin chirality obtained in the MFA for several commensurate magnetic orders. 
In Sect. 3.2, 
we show the magnon dispersion and specific heat 
obtained in the LSWA for four chiral magnets. 

\subsection{Ground-state spin configurations, energies, and spin chirality}

We start with the results of nonchiral orders in the MFA. 
Their results are summarized in Table I. 
The FM $J_{0}$ stabilizes the collinear FM order, 
and the AF $J_{0}$ stabilizes the collinear AF order and coplanar AF orders. 
Also, 
the collinear FM and AF orders 
are independent of the other exchange interactions, 
and the coplanar AF orders depend on $D_{0}$. 
Because of this property, 
the spin vector chirality is finite only in the coplanar AF orders. 
Even in the coplanar AF orders, 
the spin scalar chirality is zero  
because 
one of the three components of 
$\langle \hat{S}_{\boldi}^{\alpha}\rangle$ is zero. 
Since $\langle \hat{\boldS}_{\boldone}\times \hat{\boldS}_{\boldtwo}\rangle$ 
and $\langle \hat{\boldS}_{\boldthree}\times \hat{\boldS}_{\boldfour}\rangle$ 
are zero in the coplanar AF orders, 
the sign change of $D_{0}$ in a coplanar AF order 
leads to both interchanging of the spin directions at sublattices 3 and 4 
and changing of the sign of $D_{0}$ in the ground-state energy. 

\begin{table*}[tb]
\vspace{-15pt}
\caption{Ground-state spin configurations, 
energies per tetrahedron, 
and spin scalar chirality in the MFA 
for the AIAO, 2I2O, and 3I1O chiral magnets. 
$S_{3}$ is given by $3(S_{3})^{2}=S^{2}=\frac{1}{4}$. 
$\langle \hat{\boldS}_{\boldr_{2}}\cdot 
(\hat{\boldS}_{\boldr_{3}}\times \hat{\boldS}_{\boldr_{4}})\rangle$ 
is the spin scalar chirality of three spins in a kagome layer, 
while $\langle \hat{\boldS}_{\boldr_{1}}\cdot 
(\hat{\boldS}_{\boldr_{2}}\times \hat{\boldS}_{\boldr_{3}})\rangle$ 
is the spin scalar chirality of 
a spin in a triangle layer and two spins in a kagome layer. 
The AIAO and 3I1O chiral magnets satisfy
$\langle \hat{\boldS}_{\boldr_{1}}\cdot 
(\hat{\boldS}_{\boldr_{2}}\times \hat{\boldS}_{\boldr_{3}})\rangle
=\langle \hat{\boldS}_{\boldr_{1}}\cdot 
(\hat{\boldS}_{\boldr_{4}}\times \hat{\boldS}_{\boldr_{2}})\rangle
=\langle \hat{\boldS}_{\boldr_{1}}\cdot 
(\hat{\boldS}_{\boldr_{3}}\times \hat{\boldS}_{\boldr_{4}})\rangle$, 
and the 2I1O chiral magnet satisfies
$\langle \hat{\boldS}_{\boldr_{1}}\cdot 
(\hat{\boldS}_{\boldr_{2}}\times \hat{\boldS}_{\boldr_{3}})\rangle
=\langle \hat{\boldS}_{\boldr_{1}}\cdot 
(\hat{\boldS}_{\boldr_{4}}\times \hat{\boldS}_{\boldr_{2}})\rangle
=-\langle \hat{\boldS}_{\boldr_{1}}\cdot 
(\hat{\boldS}_{\boldr_{3}}\times \hat{\boldS}_{\boldr_{4}})\rangle$. 
}
\label{tab:chiral}
\vspace{5pt}
\begin{tabular}{ccccccccc} \hline\\[-10pt]
\multicolumn{1}{c}{ \ \ \ \ \ \ \ \ \ \ } &
\multicolumn{1}{c}{ \ \ AIAO  \ \ } & 
\multicolumn{1}{c}{ \ \ 2I2O \ \ } &
\multicolumn{1}{c}{ \ \ 3I1O \ \ } \\[1pt] \hline\\[-5pt]
$\langle \hat{\boldS}_{\boldr_{1}}\rangle$ 
&${}^{t}(S_{3}\ S_{3}\ S_{3})$ 
&${}^{t}(S_{3}\ S_{3}\ S_{3})$ 
&${}^{t}(S_{3}\ S_{3}\ S_{3})$\\[5pt]
$\langle \hat{\boldS}_{\boldr_{2}}\rangle$ 
&${}^{t}(-S_{3}\ -S_{3}\ S_{3})$ 
&${}^{t}(-S_{3}\ -S_{3}\ S_{3})$ 
&${}^{t}(S_{3}\ S_{3}\ -S_{3})$\\[5pt]
$\langle \hat{\boldS}_{\boldr_{3}}\rangle$ 
&${}^{t}(S_{3}\ -S_{3}\ -S_{3})$ 
&${}^{t}(-S_{3}\ S_{3}\ S_{3})$ 
&${}^{t}(-S_{3}\ S_{3}\ S_{3})$\\[5pt]
$\langle \hat{\boldS}_{\boldr_{4}}\rangle$ 
&${}^{t}(-S_{3}\ S_{3}\ -S_{3})$ 
&${}^{t}(S_{3}\ -S_{3}\ S_{3})$ 
&${}^{t}(S_{3}\ -S_{3}\ S_{3})$\\[5pt]\hline \\[-10pt] 
$\frac{\langle \hat{H}_{\textrm{eff}}\rangle}{N}$ 
&$-J_{0}-4D_{0}$ 
&$\dfrac{1}{3}J_{0}+\dfrac{4}{3}D_{0}$
&$J_{1}+4D_{1}$ \\[5pt]\hline \\[-7pt]
$\langle \hat{\boldS}_{\boldr_{2}}\cdot 
(\hat{\boldS}_{\boldr_{3}}\times \hat{\boldS}_{\boldr_{4}})\rangle$ 
&$-\frac{1}{6\sqrt{3}}$ 
&$-\frac{1}{6\sqrt{3}}$ 
&$\frac{1}{6\sqrt{3}}$\\[8pt]
$\langle \hat{\boldS}_{\boldr_{1}}\cdot 
(\hat{\boldS}_{\boldr_{2}}\times \hat{\boldS}_{\boldr_{3}})\rangle$ 
&$\frac{1}{6\sqrt{3}}$ 
&$-\frac{1}{6\sqrt{3}}$ 
&$\frac{1}{6\sqrt{3}}$\\[8pt]\hline \\[-8pt] 
\end{tabular}
\vspace{-10pt}
\end{table*}

\begin{figure}[tb]
\includegraphics[width=67mm]{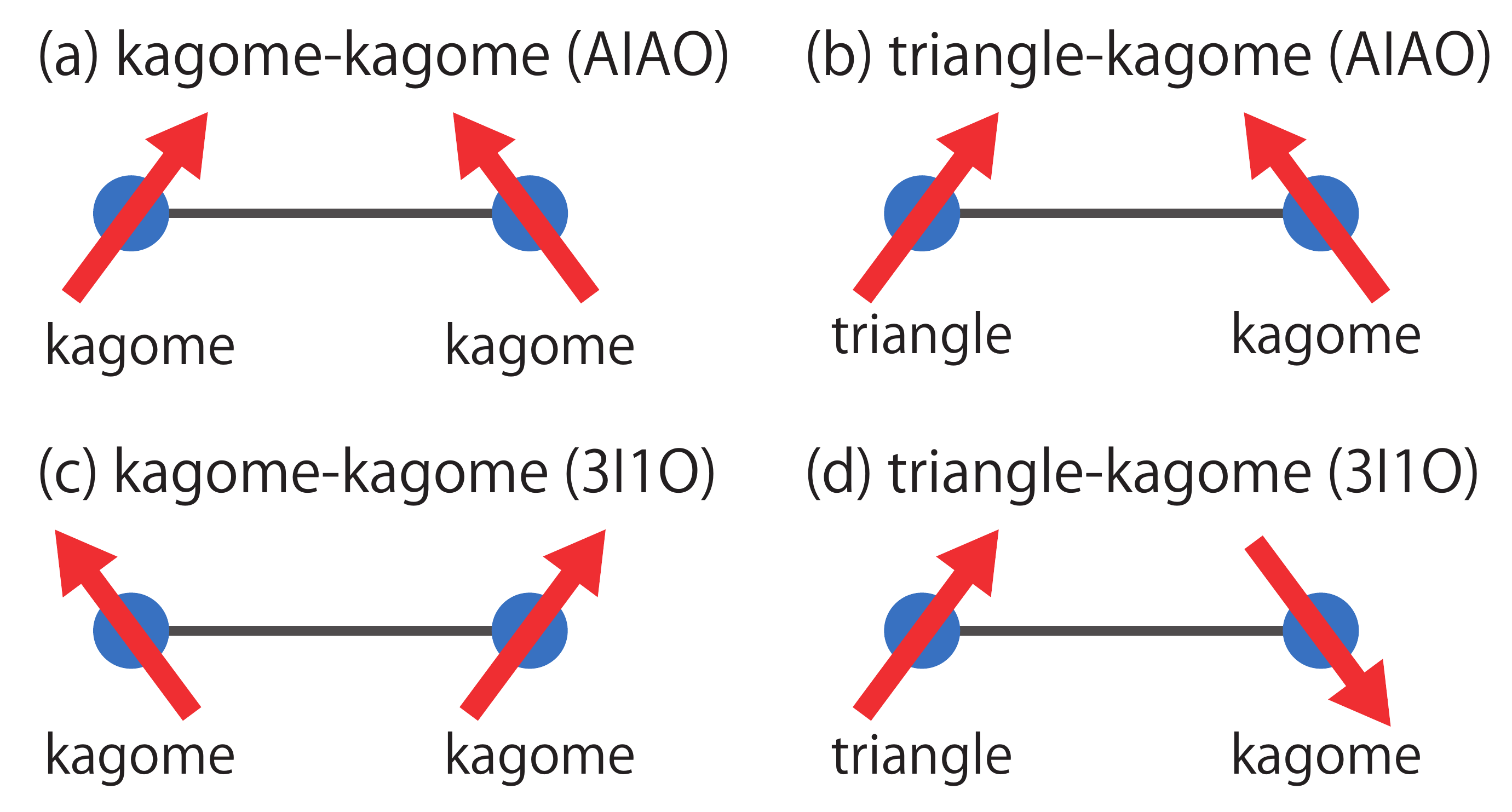}
\vspace{-15pt}
\caption{
(Color online)
Twisted spins of the NN bonds 
in (a), (b) the AIAO chiral order 
and (c), (d) the 3I1O chiral order. 
}
\label{fig4}
\end{figure}

We next show the results for typical chiral orders of pyrochlore magnets in the MFA. 
We summarize the results in Table II. 

First, 
the AF $J_{0}$ and positive $D_{0}$ stabilize 
the AIAO chiral order. 
This is because the AIAO chiral order 
possesses a spin structure  
in which 
the total Heisenberg interactions between NN bonds are AF 
and the spins for all the NN bonds are twisted in the same way, 
as shown in Figs. \ref{fig4}(a) and \ref{fig4}(b). 
This chiral order becomes the most stable ground state, for example, 
at $J_{0}=1$, $D_{0}=0.2$, and $J_{1}=D_{1}=0$. 
In this chiral order, 
the spin scalar chirality of three spins 
in a kagome layer is the same in magnitude 
as that of a spin in a triangle layer and two spins in a kagome layer, 
while the signs are opposite. 

Second, 
the FM $J_{0}$ and positive $D_{0}$ stabilize 
the 2I2O chiral order. 
In contrast to the AIAO chiral order, 
this chiral order is not the most stable 
in our model
because the energy reduction of the $J_{0}$ and $D_{0}$ terms 
is not large. 

Third, 
the combination of the FM $J_{1}$ and negative $D_{1}$ stabilizes 
the 3I1O chiral order. 
This is because 
the 3I1O chiral order 
possesses the spin structure in which 
the total Heisenberg interactions between 
the NN bonds of two spins in a kagome layer are AF 
and the spins are twisted in the way shown in Fig. \ref{fig4}(c), 
while the total Heisenberg interactions between 
the NN bonds of a spin in a triangle layer 
and a spin in a kagome layer are FM 
and the spins are twisted in the way shown in Fig. \ref{fig4}(d). 
(This difference 
is the key to understanding the kinds of branches 
of the magnon dispersion, as we will explain in Sect. 3.2.1.)
In the 3I1O chiral order, 
the spin scalar chirality of three spins in a kagome layer 
are the same in both magnitude and sign 
as the spin scalar chirality of a spin in a triangle layer 
and two spins in a kagome layer. 
This differs from the property of the AIAO and 2I2O chiral orders. 

\begin{table*}[tb]
\vspace{-15pt}
\caption{Ground-state 
spin configurations, energies per tetrahedron, 
and spin scalar chirality in the MFA for the dAIAO and d3I1O chiral magnets.
$S_{3}$ is given by $3(S_{3})^{2}=S^{2}=\frac{1}{4}$, 
and $S_{3}^{\prime}$ and $S^{\prime\prime}_{3}$ are given by 
$2(S^{\prime}_{3})^{2}+(S^{\prime\prime}_{3})^{2}=S^{2}=\frac{1}{4}$. 
$\langle \hat{\boldS}_{\boldr_{2}}\cdot 
(\hat{\boldS}_{\boldr_{3}}\times \hat{\boldS}_{\boldr_{4}})\rangle$ 
is the spin scalar chirality of three spins in a kagome layer, 
while $\langle \hat{\boldS}_{\boldr_{1}}\cdot 
(\hat{\boldS}_{\boldr_{2}}\times \hat{\boldS}_{\boldr_{3}})\rangle$ 
is the spin scalar chirality of 
a spin in a triangle layer and two spins in a kagome layer. 
These chiral magnets satisfy
$\langle \hat{\boldS}_{\boldr_{1}}\cdot 
(\hat{\boldS}_{\boldr_{2}}\times \hat{\boldS}_{\boldr_{3}})\rangle
=\langle \hat{\boldS}_{\boldr_{1}}\cdot 
(\hat{\boldS}_{\boldr_{4}}\times \hat{\boldS}_{\boldr_{2}})\rangle
=\langle \hat{\boldS}_{\boldr_{1}}\cdot 
(\hat{\boldS}_{\boldr_{3}}\times \hat{\boldS}_{\boldr_{4}})\rangle$. 
}
\label{tab:dischiral}
\vspace{5pt}
\begin{tabular}{ccccccccc} \hline\\[-10pt]
\multicolumn{1}{c}{ \ \ \ \ \ \ \ \ \ \ } &
\multicolumn{1}{c}{ \ \ \ \ \ \ \ dAIAO \ \ \ \ \ } & 
\multicolumn{1}{c}{ \ \ \ \ \ \ \ d3I1O \ \ \ \ \ } \\[1pt] \hline\\[-5pt]
$\langle \hat{\boldS}_{\boldr_{1}}\rangle$ 
&${}^{t}(S_{3}\ S_{3}\ S_{3})$ 
&${}^{t}(S_{3}\ S_{3}\ S_{3})$
\\[5pt]
$\langle \hat{\boldS}_{\boldr_{2}}\rangle$ 
&${}^{t}(-S_{3}^{\prime}\ -S_{3}^{\prime}\ S_{3}^{\prime\prime})$
&${}^{t}(S_{3}^{\prime}\ S_{3}^{\prime}\ -S_{3}^{\prime\prime})$
\\[5pt]
$\langle \hat{\boldS}_{\boldr_{3}}\rangle$ 
&${}^{t}(S_{3}^{\prime\prime}\ -S_{3}^{\prime}\ -S_{3}^{\prime})$
&${}^{t}(-S_{3}^{\prime\prime}\ S_{3}^{\prime}\ S_{3}^{\prime})$
\\[5pt]
$\langle \hat{\boldS}_{\boldr_{4}}\rangle$ 
&${}^{t}(-S_{3}^{\prime}\ S_{3}^{\prime\prime}\ -S_{3}^{\prime})$
&${}^{t}(S_{3}^{\prime}\ -S_{3}^{\prime\prime}\ S_{3}^{\prime})$
\\[5pt]\hline \\[-10pt] 
$\frac{\langle \hat{H}_{\textrm{eff}}\rangle}{N}$ 
&$6(J_{0}+J_{1})S_{3}(S_{3}^{\prime\prime}-2S^{\prime}_{3})$ 
&$6(J_{0}+J_{1})S_{3}(2S^{\prime}_{3}-S_{3}^{\prime\prime})$\\[5pt]
\
&$+6(J_{0}-J_{1})S_{3}^{\prime}(S_{3}^{\prime}-2S^{\prime\prime}_{3})$ 
&$+6(J_{0}-J_{1})S_{3}^{\prime}(S_{3}^{\prime}-2S^{\prime\prime}_{3})$\\[5pt]
\  
&$
-12(D_{0}+D_{1})S_{3}(S_{3}^{\prime\prime}+S^{\prime}_{3})$ 
&$
+12(D_{0}+D_{1})S_{3}(S_{3}^{\prime\prime}+S^{\prime}_{3})$\\[5pt]
\  
&$
-12(D_{0}-D_{1})S_{3}^{\prime}(S_{3}^{\prime\prime}+S^{\prime}_{3})$ 
&$
-12(D_{0}-D_{1})S_{3}^{\prime}(S_{3}^{\prime\prime}+S^{\prime}_{3})$\\[5pt]\hline \\[-7pt]
$\langle \hat{\boldS}_{\boldr_{2}}\cdot 
(\hat{\boldS}_{\boldr_{3}}\times \hat{\boldS}_{\boldr_{4}})\rangle$ 
&$(S^{\prime}_{3}+S_{3}^{\prime\prime})[S^{\prime\prime}_{3}(S_{3}^{\prime\prime}-S^{\prime}_{3})-2(S_{3}^{\prime})^{2}]$ 
&$-(S^{\prime}_{3}+S^{\prime\prime}_{3})[S^{\prime\prime}_{3}(S^{\prime\prime}_{3}-S^{\prime}_{3})-2(S^{\prime}_{3})^{2}]$\\[8pt]
$\langle \hat{\boldS}_{\boldr_{1}}\cdot 
(\hat{\boldS}_{\boldr_{2}}\times \hat{\boldS}_{\boldr_{3}})\rangle$ 
&$S_{3}(S^{\prime}_{3}+S_{3}^{\prime\prime})^{2}$ 
&$S_{3}(S^{\prime}_{3}+S_{3}^{\prime\prime})^{2}$\\[6pt]\hline \\[-8pt] 
\end{tabular}
\vspace{-10pt}
\end{table*}

Then, 
there are other chiral orders, whose ground-state energies depend on 
$J_{0}$, $D_{0}$, $J_{1}$, and $D_{1}$. 
Their results are summarized in Table III. 

First, 
the dAIAO chiral order 
is stabilized by the combination of 
the AF $J_{0}$, the positive $D_{0}$, and the small $J_{1}$ and $D_{1}$. 
In this chiral order, 
the directions of the spins in a tetrahedron 
intersect at a point 
on the line connecting sublattice 1 and the center of the tetrahedron.   
Since in the AIAO chiral order 
the directions of the spins in a tetrahedron 
intersect at the center of the tetrahedron, 
the difference between these chiral orders 
is the displacement of the intersecting point. 
We can thus regard this 
as a distorted AIAO chiral order. 
As we see from Table III, 
the dAIAO chiral order possesses two properties 
that are different from the properties of the AIAO chiral order: 
one is the dependence of the ground-state energy on $J_{1}$ and $D_{1}$; 
the other is the bond-dependent magnitude of the spin scalar chirality. 
The former originates from 
the combination of the stabilization of the AIAO-like spin structure 
due to $J_{0}$ and $D_{0}$ and 
the modification of the spin structure 
due to the corrections of $J_{1}$ and $D_{1}$; 
the latter originates from the imbalance between 
the exchange interactions in a kagome layer 
and those between triangle and kagome layers 
due to the corrections of $J_{1}$ and $D_{1}$. 
Actually, 
we can more clearly understand the origins of the two properties 
by analyzing the slightly distorted case.
In this case, 
the spin structure in a tetrahedron 
is determined by choosing 
$S^{\prime}_{3}$ and $S^{\prime\prime}_{3}$ 
in the spin structure of the dAIAO chiral order 
as $S^{\prime}_{3}=S_{3}+\Delta S^{\prime}$ 
and $S^{\prime\prime}_{3}=S_{3}-2\Delta S^{\prime}$, 
where $\Delta S^{\prime}$ is a small correction; 
$\Delta S^{\prime}$ has been chosen 
because the $S_{3}^{\prime}$ and $S^{\prime\prime}_{3}$ 
in this choice satisfy the hard-spin constraint within the $O(\Delta S^{\prime})$ terms.  
By substituting 
$S^{\prime}_{3}=S_{3}+\Delta S^{\prime}$ 
and $S^{\prime\prime}_{3}=S_{3}-2\Delta S^{\prime}$ 
in $\frac{\langle \hat{H}_{\textrm{eff}}\rangle}{N}$, 
$\langle \hat{\boldS}_{\boldr_{2}}\cdot 
(\hat{\boldS}_{\boldr_{3}}\times \hat{\boldS}_{\boldr_{4}})\rangle$, 
and $\langle \hat{\boldS}_{\boldr_{1}}\cdot 
(\hat{\boldS}_{\boldr_{2}}\times \hat{\boldS}_{\boldr_{3}})\rangle$ 
of the dAIAO chiral order for Table III 
and estimating them within the $O(\Delta S^{\prime})$ terms, 
we obtain 
\begin{align}
\frac{\langle \hat{H}_{\textrm{eff}}\rangle}{N}
=(-J_{0}-4D_{0})+4\sqrt{3} (-2J_{1}+D_{1})\Delta S^{\prime},\label{eq:E_AIAO-simple}
\end{align}
and 
\begin{align}
\langle \hat{\boldS}_{\boldr_{2}}\cdot 
(\hat{\boldS}_{\boldr_{3}}\times \hat{\boldS}_{\boldr_{4}})\rangle
=&-\frac{1}{6\sqrt{3}}-\Delta S^{\prime},\label{eq:SSS-1_AIAO-simple}\\
\langle \hat{\boldS}_{\boldr_{1}}\cdot 
(\hat{\boldS}_{\boldr_{3}}\times \hat{\boldS}_{\boldr_{4}})\rangle
=&\frac{1}{6\sqrt{3}}-\frac{1}{3}\Delta S^{\prime}.\label{eq:SSS-2_AIAO-simple}
\end{align}

Second, 
the d3I1O chiral order 
is stabilized by the combination of 
the FM $J_{1}$, the negative $D_{1}$, 
and the corrections of $J_{0}$ and $D_{0}$. 
We can understand its stabilizing mechanism and properties, which are different from 
the properties of the 3I1O chiral order, 
in a similar way as for the dAIAO chiral order. 
This chiral order 
can be regarded as a distorted 3I1O chiral order 
because the difference between it and 
the 3I1O chiral order 
is the displacement of the intersection point of the directions of the spins 
in a tetrahedron. 
The d3I1O chiral order 
is stabilized 
by the combination 
of the stabilization of a 3I1O-like spin structure 
due to $J_{1}$ and $D_{1}$ 
and the modification of the spin structure 
due to the corrections of $J_{0}$ and $D_{0}$. 
As a result, 
its ground-state energy depends on not only $J_{1}$ and $D_{1}$ 
but also $J_{0}$ and $D_{0}$. 
Furthermore, 
there is another property different from the property of 
the 3I1O chiral order: 
the bond-dependent magnitude of the spin scalar chirality 
due to the imbalance between the exchange interactions in a kagome layer 
and the exchange interactions between triangle and kagome layers. 
The origins of these two properties can be more clearly understood 
by analyzing a slightly distorted case of the d3I1O 
chiral order: 
we first choose 
$S^{\prime}_{3}$ and $S^{\prime\prime}_{3}$ 
in the spin structure of the d3I1O chiral order 
as $S^{\prime}_{3}=S_{3}+\Delta S^{\prime}$ and $S^{\prime\prime}_{3}=S_{3}-2\Delta S^{\prime}$ 
with $\Delta S^{\prime}$ 
because $S_{3}^{\prime}$ and $S_{3}^{\prime\prime}$ satisfy 
the hard-spin constraint within the $O(\Delta S^{\prime})$ terms; 
we then
estimate the ground-state energy and spin scalar chirality
within the $O(\Delta S^{\prime})$ terms as
\begin{align}
\frac{\langle \hat{H}_{\textrm{eff}}\rangle}{N}
=(J_{1}+4D_{1})-4\sqrt{3}(-2J_{0}+D_{0})\Delta S^{\prime},\label{eq:E_3I1O-simple}
\end{align}
and 
\begin{align}
\langle \hat{\boldS}_{\boldr_{2}}\cdot 
(\hat{\boldS}_{\boldr_{3}}\times \hat{\boldS}_{\boldr_{4}})\rangle
=&\frac{1}{6\sqrt{3}}+\Delta S^{\prime},\label{eq:SSS-1_3IA1-simple}\\
\langle \hat{\boldS}_{\boldr_{1}}\cdot 
(\hat{\boldS}_{\boldr_{3}}\times \hat{\boldS}_{\boldr_{4}})\rangle
=&\frac{1}{6\sqrt{3}}-\frac{1}{3}\Delta S^{\prime}.\label{eq:SSS-2_3I1O-simple}
\end{align}

\begin{figure*}[tb]
\includegraphics[width=136mm]{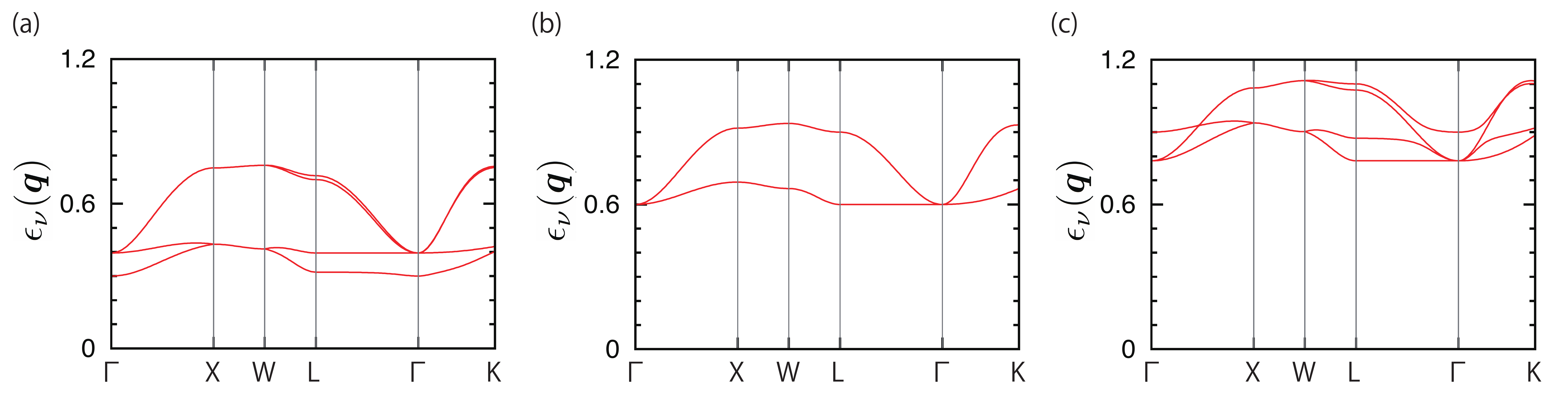}
\vspace{-14pt}
\caption{
(Color online)
Magnon dispersion curves, $\epsilon_{\nu}(\boldq)$ ($\nu=1,\cdots,4$), 
in the LSWA for
the AIAO chiral magnet 
at $J_{0}=1$, $J_{1}=D_{1}=0$, 
and $D_{0}=$ (a) $0.1$, (b) $0.2$, and (c) $0.3$. 
The symmetrical points are  
$\Gamma(0\ 0\ 0)$, X$(\pi \ 0\ 0)$, W$(\pi \ \frac{\pi}{2}\ 0)$, 
L$(\frac{\pi}{2} \ \frac{\pi}{2} \ \frac{\pi}{2})$, 
and K$(\frac{3\pi}{4} \ \frac{3\pi}{4} \ 0)$. 
}
\label{fig6}
\end{figure*}
\begin{figure*}[tb]
\includegraphics[width=136mm]{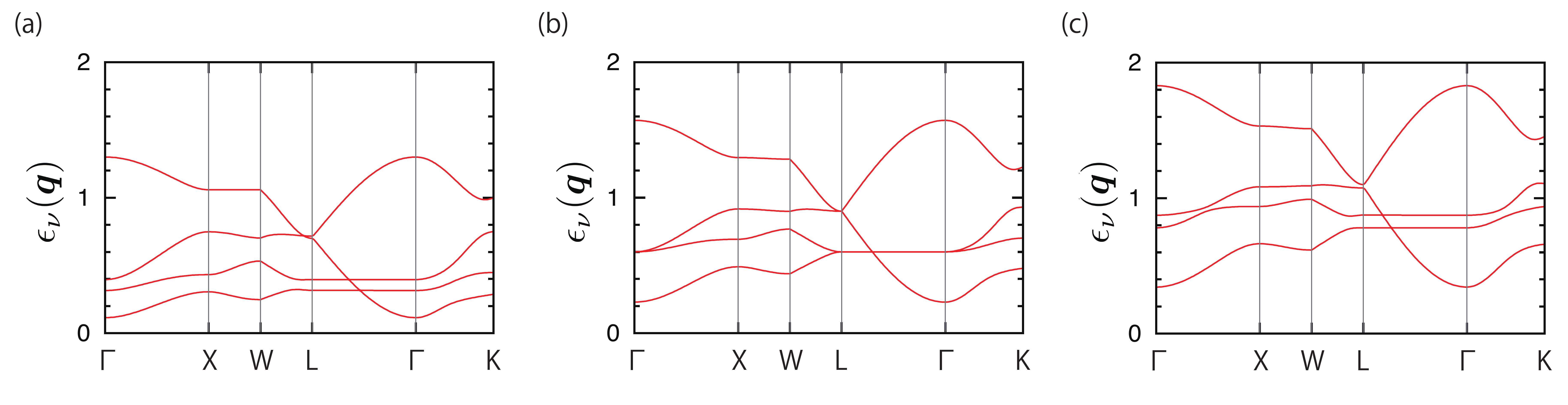}
\vspace{-14pt}
\caption{
(Color online)
Magnon dispersion curves, $\epsilon_{\nu}(\boldq)$ ($\nu=1,\cdots,4$), 
in the LSWA for
the 3I1O chiral magnet 
at $J_{1}=-1$, $J_{0}=D_{0}=0$, 
and $D_{1}=$ (a) $-0.1$, (b) $-0.2$, and (c) $-0.3$. 
}
\label{fig7}
\end{figure*}
\subsection{Magnon dispersion and specific heat}

We turn to the magnon dispersion and specific heat for four chiral magnets 
in our $S=\frac{1}{2}$ model in the LSWA. 
The four chiral magnets are the AIAO, 
3I1O, dAIAO, and d3I1O chiral magnets; 
their ordering vectors are all $\boldQ=\boldzero$.  
We consider these magnets because 
one of them can be the most stable ground state 
of our $S=\frac{1}{2}$ model in the MFA, 
depending on the values of $J_{0}$, $D_{0}$, $J_{1}$, and $D_{1}$. 
We show the results for the magnon dispersion and specific heat 
in Sects. 3.2.1 and 3.2.2, respectively. 

We obtained the results by numerical calculations. 
In the numerical calculations, 
we substituted $N_{x}=N_{y}=N_{z}=128$ 
in $q_{x}=\pi(\frac{m_{x}}{N_{x}}-\frac{m_{y}}{N_{y}}+\frac{m_{z}}{N_{z}})$,
$q_{y}=\pi(\frac{m_{x}}{N_{x}}+\frac{m_{y}}{N_{y}}-\frac{m_{z}}{N_{z}})$, and 
$q_{z}=\pi(-\frac{m_{x}}{N_{x}}+\frac{m_{y}}{N_{y}}+\frac{m_{z}}{N_{z}})$,
with $N_{x}N_{y}N_{z}=N$, $1\le m_{x}\le N_{x}$, $1\le m_{y}\le N_{y}$, and $1\le m_{z}\le N_{z}$; 
we chose the convergence factor $\Delta$ 
as $\Delta=10^{-7}$, 
which is tiny compared with the finite exchange interactions; 
we chose the values of $J_{0}$, $D_{0}$, $J_{1}$, and $D_{1}$ 
so as to enable one of the four chiral orders to be most stable 
at zero temperature in the MFA. 
For the AIAO chiral magnet, 
we set $J_{0}=1$, $D_{0}=0.1$, $0.2$, or $0.3$, 
and $J_{1}=D_{1}=0$. 
Since the corrections of $J_{1}$ and $D_{1}$ to the AIAO chiral order 
cause the dAIAO chiral order to be more stable, 
we set $J_{0}=1$, $D_{0}=0.2$ or $0.3$, 
$J_{1}=0.5$ or $-0.5$, 
and $D_{1}=\frac{J_{1}D_{0}}{|J_{0}|}=J_{1}D_{0}$ 
for the dAIAO chiral magnet. 
Then, 
we set $J_{0}=D_{0}=0$, $J_{1}=-1$, and $D_{1}=-0.1$, $-0.2$, or $-0.3$ 
for the 3I1O chiral magnet. 
For the d3I1O chiral magnet, 
we set $J_{0}=-1$, $J_{1}=-1.8$, $D_{0}=0.1$ or $0.2$, 
and $D_{1}=\frac{J_{1}D_{0}}{|J_{0}|}=J_{1}D_{0}$. 
Only for the 3I1O chiral magnet, 
the magnitude of $J_{1}$ is chosen as the energy unit; 
for the others, 
the energy unit is the magnitude of $J_{0}$. 
For each set of the parameters, 
we numerically calculated the noninteracting magnon dispersion of a chiral magnet 
using the algorithm explained in Sect. 2.3. 
We also calculated the specific heat, $C_{v}$, given by 
\begin{align}
C_{v}
=&
\frac{2}{T^{2}}
\sum\limits_{\boldq}
\sum\limits_{\nu=1}^{4}
\epsilon_{\nu}(\boldq)^{2}
n[\epsilon_{\nu}(\boldq)]^{2}
\exp[\frac{\epsilon_{\nu}(\boldq)}{T}],\label{eq:Cv}
\end{align} 
where $n(\epsilon)$ is the Bose distribution function.

\subsubsection{Magnon dispersion}
\begin{figure}[tb]
\includegraphics[width=82mm]{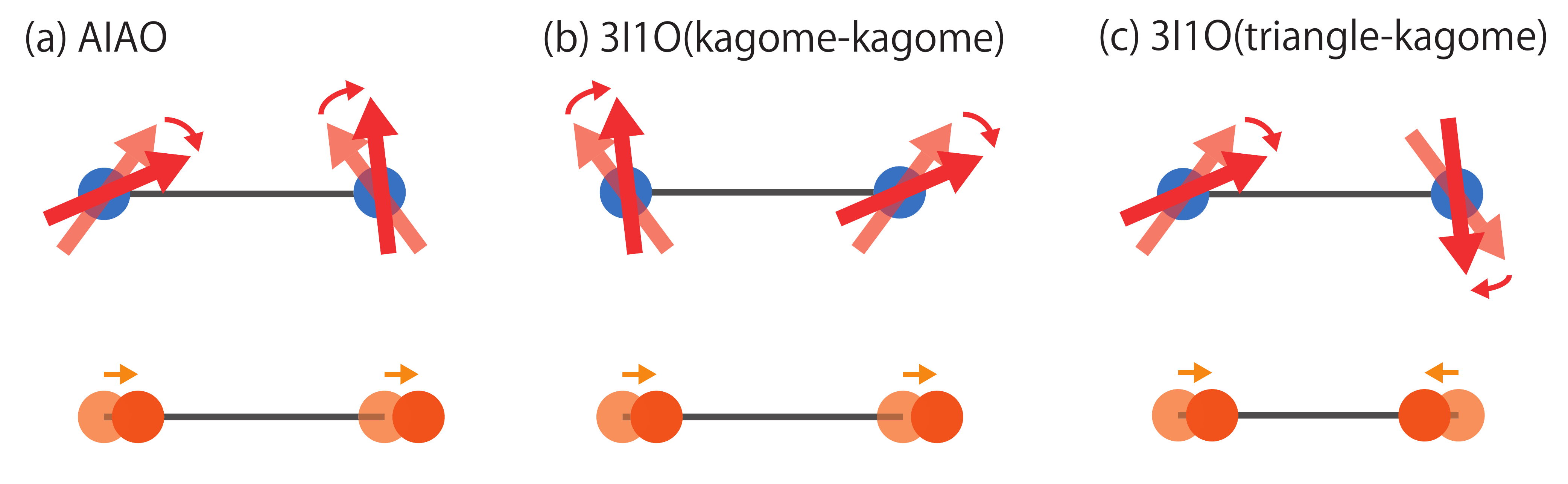}
\caption{
(Color online)
Rotations of two spins (upper rows) and the analogous motion of phonons (lower rows)
in (a) the AIAO chiral magnet 
and (b), (c) the 3I1O chiral magnet. }
\label{fig8}
\end{figure}
\begin{figure}[tb]
{\hspace{-10pt}
\includegraphics[width=74mm]{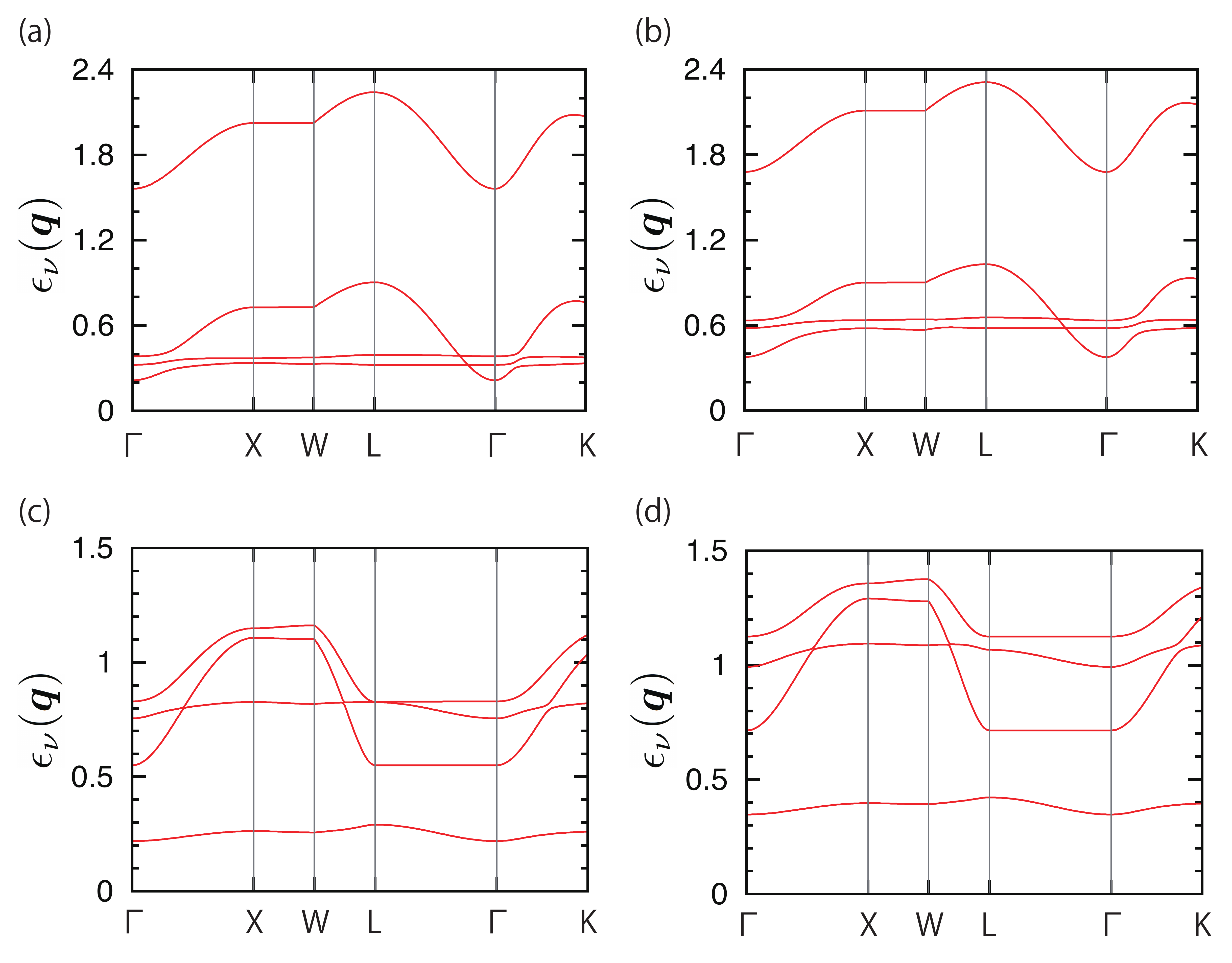}}
\vspace{-14pt}
\caption{
(Color online)
Magnon dispersion curves, $\epsilon_{\nu}(\boldq)$ ($\nu=1,\cdots,4$), 
in the LSWA for
the dAIAO chiral magnet 
at 
$J_{0}=1$, $D_{1}=J_{1}D_{0}$, and 
$(J_{1},D_{0})=$ (a) $(0.5,0.2)$, (b) $(0.5,0.3)$, 
(c) $(-0.5,0.2)$, and (d) $(-0.5,0.3)$. 
}
\label{fig9}
\end{figure}
\begin{figure}[tb]
{\hspace{-10pt}
\includegraphics[width=74mm]{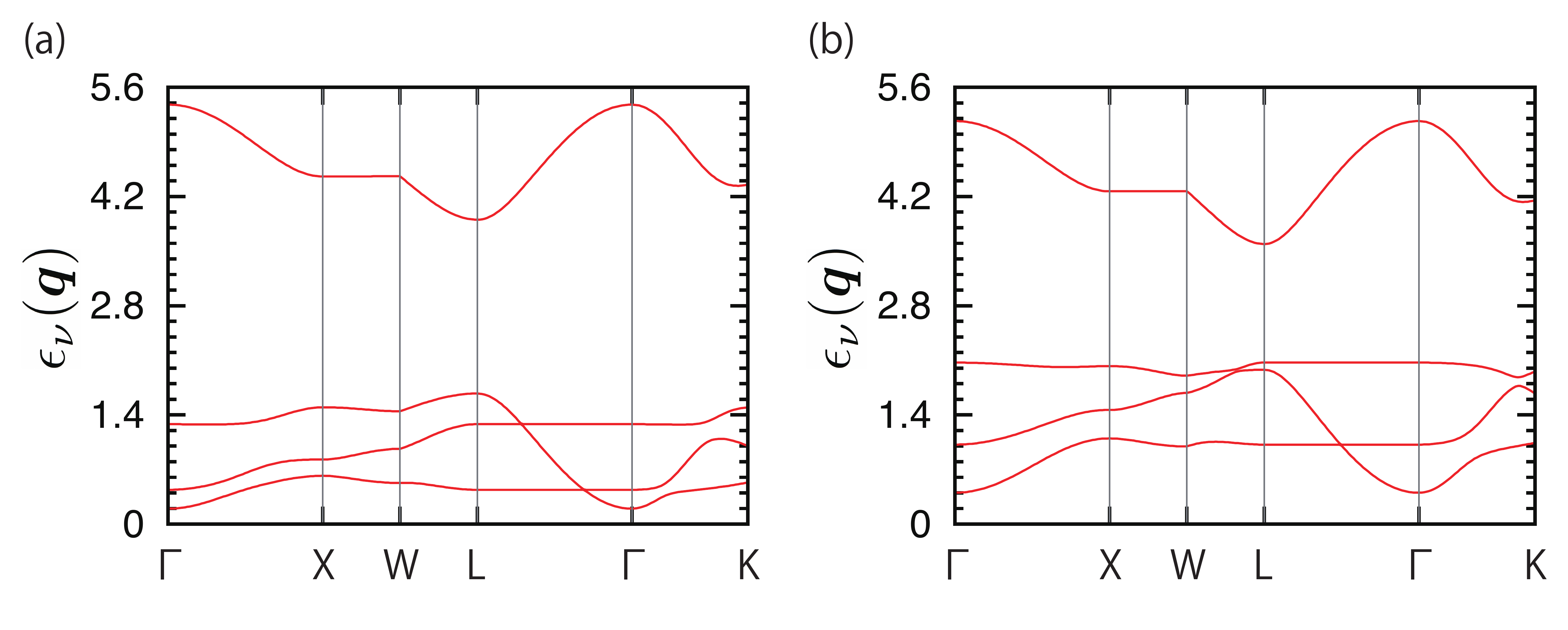}}
\vspace{-14pt}
\caption{
(Color online)
Magnon dispersion curves, $\epsilon_{\nu}(\boldq)$ ($\nu=1,\cdots,4$), 
in the LSWA for
the d3I1O chiral magnet 
at $J_{0}=-1$, $J_{1}=-1.8$, $D_{1}=J_{1}D_{0}$,   
and $D_{0}=$ (a) $0.2$, and (b) $0.3$. 
}
\label{fig10}
\end{figure}

Before giving the results for the magnon dispersions, 
we define a quasiacoustic or an optical branch 
and explain the appropriateness of the definitions. 
A quasiacoustic branch is defined as a branch that increases 
with increasing displacement from $\boldq=\boldQ$, 
while an optical branch is defined as a branch that decreases. 
In other words, 
the quasiacoustic and optical branches are respectively defined as 
convex and concave functions of $\boldq$ around $\boldq=\boldQ$. 
(The convex or concave function means that 
the energy at $\boldq=\boldQ$ is smallest or largest, respectively, 
in momenta near $\boldq=\boldQ$.) 
We do not impose any conditions on the value at $\boldq=\boldQ$ or 
the power of the $\boldq$ dependence. 
This is because the above definitions are appropriate for distinguishing 
whether the mode of a branch is acoustic type or optical type. 
The acoustic type mode corresponds to the same-phase motion, 
while the optical type mode corresponds to 
the $180^{\circ}$-out-of-phase motion~\cite{Ashcroft-Mermin}. 
Since the same-phase and $180^{\circ}$-out-of-phase motion 
have the lowest and largest energies, respectively, 
for $\boldq=\boldQ$, 
the above definitions are appropriate. 
We use the quasiacoustic mode rather than the acoustic mode 
because the acoustic mode in the exact sense appears 
only in a few cases, such as antiferromagnets. 

We start with the magnon dispersion of the AIAO chiral magnet. 
We can see its three properties from Fig. \ref{fig6}. 
First, 
a gap is opened at the $\Gamma$ point. 
(This gap is not due to the finite convergence factor $\Delta=10^{-7}$ 
because $\Delta$ is much smaller than the value of the gap; 
as we will see below, 
the situation is the same even in the other three chiral magnets.) 
Since this gap increases as a function of $D_{0}$ 
and is not equal to the value of $D_{0}$, 
this gap is induced by the combination of $D_{0}$ and $J_{0}$. 
This property is distinct from the magnon dispersion 
of a nonchiral magnet 
because a commensurate nonchiral magnet has gapless excitation 
at $\boldq=\boldzero$. 
Second, 
the magnon energies at the $\Gamma$ point are partially or completely degenerate: 
the degeneracy is fourfold at $D_{0}=0.2$ 
and threefold at $D_{0}=0.1$ and $0.3$. 
The partial lifting of the degeneracy arises from 
the DM interaction 
because the difference between the nondegenerate energies at the $\Gamma$ point 
is approximately equal to $D_{0}-0.2$. 
Third, 
all the branches are quasiacoustic near the $\Gamma$ point. 
This is similar for a nonchiral magnet. 

Figure \ref{fig7} shows three properties 
of the magnon dispersion of the 3I1O chiral magnet. 
First, 
a gap is opened at the $\Gamma$ point 
in a similar way to that for the AIAO chiral magnet. 
This gap is mainly induced by $D_{1}$ 
because the value of this gap approximately corresponds to the magnitude of $D_{1}$. 
Second, 
the magnon energies at the $\Gamma$ point are degenerate  
only at $D_{1}=-0.2$, 
while at $D_{1}=-0.1$ and $-0.3$ 
the degeneracy is completely lifted. 
The lifting of the twofold degeneracy arises from $D_{1}$ 
because the difference between the second and third lowest energy branches 
at the $\Gamma$ point is approximately equal to $|D_{1}+0.2|$. 
Third, 
the highest-energy branch is optical near the $\Gamma$ point,   
while the other three branches are quasiacoustic. 
The appearance of the optical branch 
is distinct from the property of 
not only a nonchiral magnet with $\boldQ=\boldzero$
but also the AIAO chiral magnet.

By a similar analysis for phonons, 
we can understand why all the branches near $\boldq=\boldzero$ 
are quasiacoustic in the AIAO chiral magnet, 
while an optical branch appears in the 3I1O chiral magnet. 
Let us begin by recalling the mechanism of the acoustic and optical branches 
for phonons~\cite{Ashcroft-Mermin}. 
For this purpose, 
we consider an one-dimensional lattice of two different ions; 
the two different ions result in a unit cell with a two-sublattice structure. 
In this case, 
the phonon dispersion has an acoustic branch and an optical branch. 
The acoustic branch arises from the same-phase displacement 
of two different ions in each unit cell 
from the equilibrium positions, 
while the optical branch arises from 
the $180^{\circ}$-out-of-phase displacement. 
Namely, 
if we understand how fluctuations modify 
the ground-state configuration determined in a classical or semiclassical theory, 
we can understand whether a branch of the dispersion 
of bosonic quasiparticles 
is quasiacoustic or optical. 
Thus, 
in a similar way to phonons, 
we can understand the mechanism of the quasiacoustic and optical branches 
for magnons. 
We first consider the AIAO chiral magnet. 
Figure \ref{fig8}(a) shows the relative configuration 
of two spins for two of the four sublattices for the AIAO chiral magnet. 
If the fluctuations cause a clockwise rotation of the left-hand-side spin 
in Fig. \ref{fig8}(a), 
the right-hand-side spin rotates as the same-phase rotation. 
Since the relative configurations of two spins are the same 
for any pair of the four sublattices, 
all the branches for the AIAO chiral magnet 
are quasiacoustic. 
Similarly, 
we can understand the type of branches for 
the 3I1O chiral magnet. 
As we see from Figs. \ref{fig8}(b) and \ref{fig8}(c), 
there are two types of relative configurations of two spins 
for two of the four sublattices 
for the 3I1O chiral magnet: 
one is for two of sublattices 2, 3, and 4, 
while the other is for sublattice 1 and sublattice 2, 3, or 4. 
In the former configuration, 
the fluctuations induce 
the same-phase rotations of the two spins [Fig. \ref{fig8}(b)], 
while in the latter configuration, 
the $180^{\circ}$-out-of-phase rotations are induced [Fig. \ref{fig8}(c)]. 
Thus,
the same-phase rotations for two 
of sublattices 2, 3, and 4 lead to 
three quasiacoustic branches of the magnon dispersion, 
and the $180^{\circ}$-out-of-phase rotations for 
sublattice 1 and sublattice 2, 3, or 4 
lead to one optical branch.

Then, 
from Figs. \ref{fig9}(a){--}\ref{fig9}(d), 
we can find three properties 
of the magnon dispersion of the dAIAO chiral magnet. 
First, 
a gap is opened at the $\Gamma$ point 
in a similar way to that for the AIAO chiral magnet. 
Second, 
in contrast to the AIAO chiral magnet, 
there is no degeneracy of the magnon energies at the $\Gamma$ point. 
Third, 
the branches near the $\Gamma$ point are all quasiacoustic. 
This is the same as the property for the AIAO chiral magnet. 
We can understand its mechanism in a similar way 
to that for the AIAO chiral magnet 
because 
the main difference between these chiral magnets 
is the modification of the directions of the spins 
at sublattices 2, 3, and 4 (see Sect. 3.1). 
Thus, 
except the absence of the degeneracy, 
the magnon dispersion of the dAIAO chiral magnet 
is similar to that of the AIAO chiral magnet.

We can also find similarities and differences 
between the 3I1O and d3I1O chiral magnets 
by comparing Figs. \ref{fig7} and \ref{fig10}. 
The main similarities 
are the gap of the lowest-energy branch 
at the $\Gamma$ point 
and the numbers of quasiacoustic and optical branches near the $\Gamma$ point; 
the main difference is 
the degeneracy of the magnon energies at the $\Gamma$ point. 
However, 
whether the degeneracy is present or absent depends on 
the value of $\frac{D_{1}}{J_{1}}$ (see Fig. \ref{fig7}). 
Thus, 
the magnon dispersion curves of the 3I1O 
and d3I1O chiral magnets are similar. 

\subsubsection{Specific heat}

\begin{figure}[tb]
\includegraphics[width=74mm]{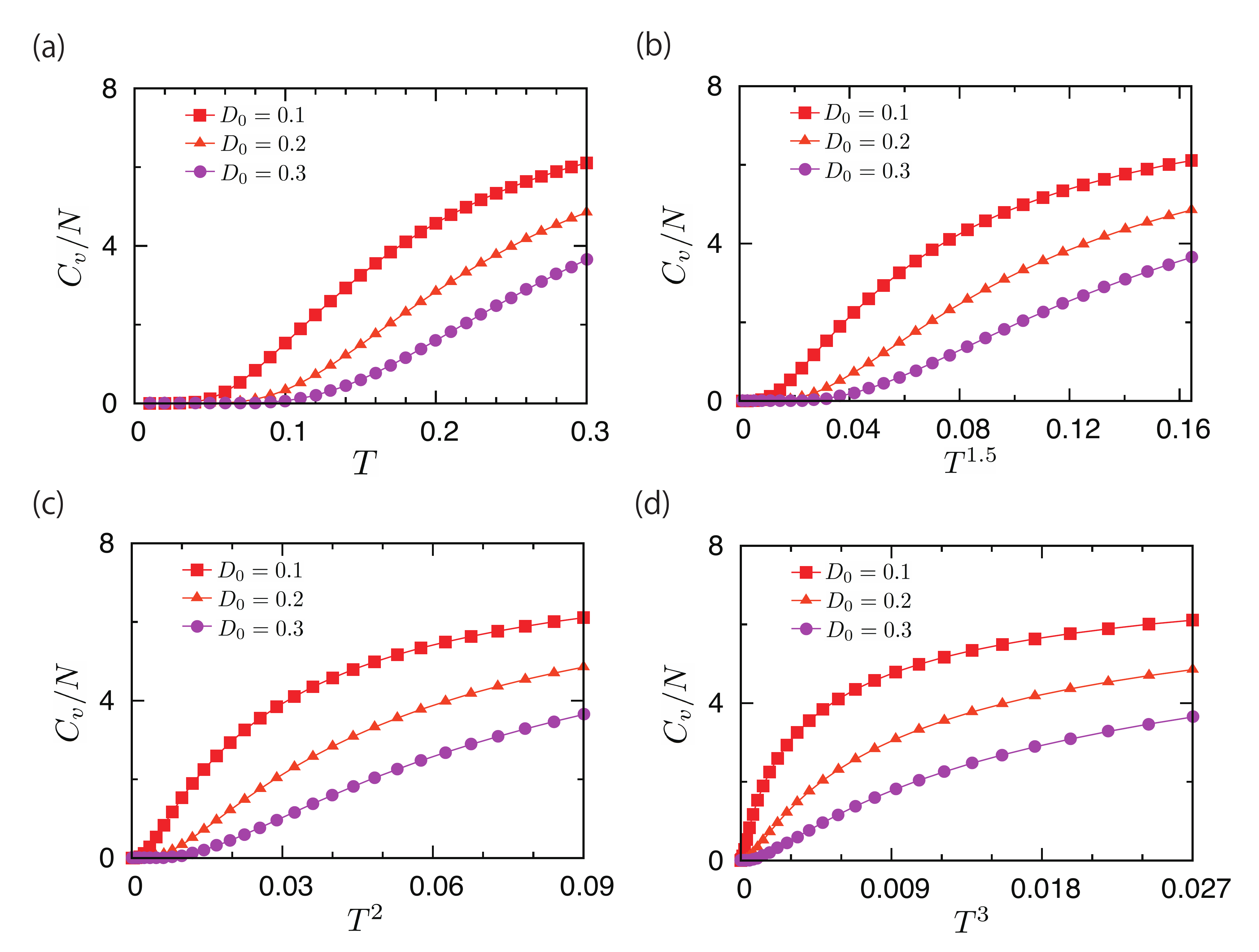}
\vspace{-15pt}
\caption{
(Color online)
Low-temperature $\frac{C_{v}}{N}$ 
in the LSWA for the AIAO chiral magnet 
as a function of (a) $T$, (b) $T^{1.5}$, (c) $T^{2}$, and (d) $T^{3}$; 
$J_{0}$, $D_{0}$, $J_{1}$, and $D_{1}$ are chosen as 
$J_{0}=1$, $J_{1}=D_{1}=0$, and $D_{0}=0.1$, $0.2$, or $0.3$. 
}
\label{fig12}
\end{figure}

\begin{figure}[tb]
\includegraphics[width=74mm]{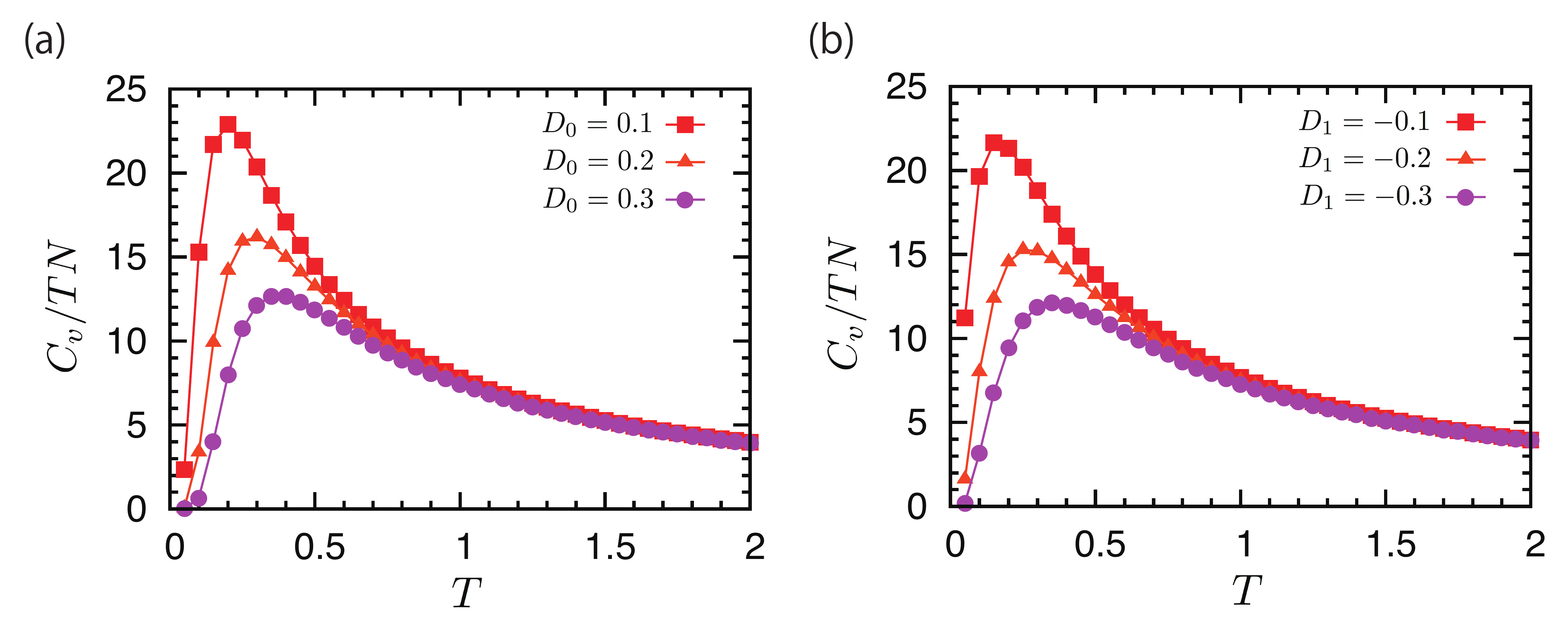}
\vspace{-15pt}
\caption{
(Color online)
Low-temperature $\frac{C_{v}}{NT}$ in the LSWA for (a) the AIAO chiral magnet 
and (b) the 3I1O chiral magnet. 
}
\label{figAdd}
\end{figure}

We turn to the results for the specific heat of the AIAO chiral magnet. 
(Since we use the LSWA, which may be appropriate only for low temperature, 
we focus on the results at low temperature.) 
The results are shown in Figs. \ref{fig12} and \ref{figAdd}(a). 
We can find three properties. 
First, 
the specific heat decreases as $D_{0}$ increases. 
This is because the dominant contributions 
to the specific heat 
are from the low-energy magnons, 
and because the increase in $D_{0}$ 
causes the increase in the lowest-energy branch 
at the $\Gamma$ point [see Figs. \ref{fig6}(a){--}\ref{fig6}(c)]. 
Second, 
it is difficult to uniquely determine the power of the temperature dependence 
of the specific heat 
in the low-temperature region. 
(The region of $0\leq T\leq 0.3$ 
can be regarded as a low-temperature region 
because the typical magnitude of $J_{0}$ is approximately $30$ meV{--}$30$ K, 
and $T=0.3=0.3J_{0}$ corresponds to $T=0.3${--}$10$ K.) 
The difficulty may arise from 
the combination of the gap in the lowest-energy magnon branch at the $\Gamma$ point 
and the complex dispersion curves. 
Third, 
the ratio of $C_{v}$ to $T$ has a peak at a low temperature 
due to the absence of gapless excitation.

\begin{figure}[tb]
\includegraphics[width=74mm]{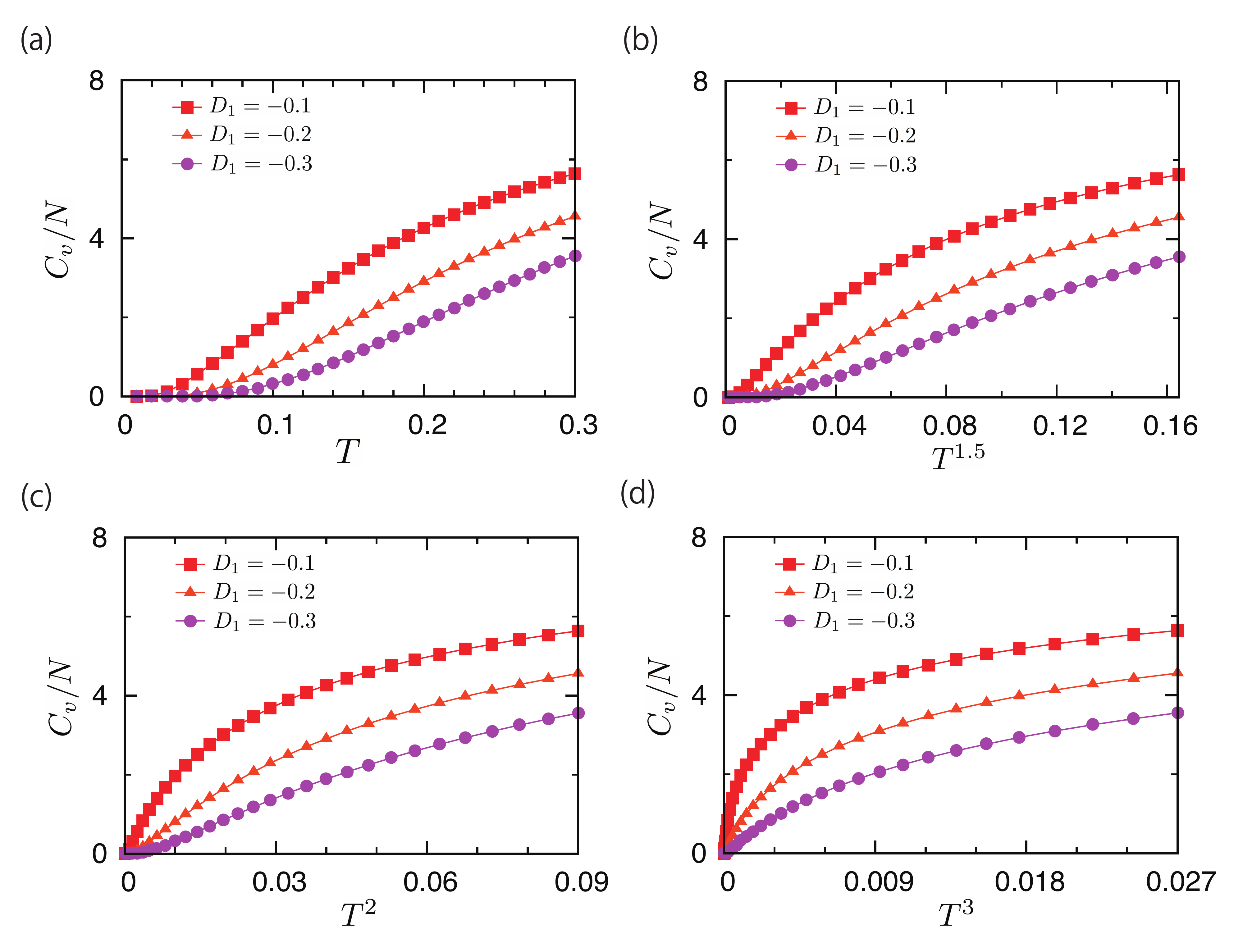}
\vspace{-15pt}
\caption{
(Color online)
Low-temperature $\frac{C_{v}}{N}$ 
in the LSWA for the 3I1O chiral magnet 
as a function of (a) $T$, (b) $T^{1.5}$, (c) $T^{2}$, and (d) $T^{3}$; 
$J_{0}$, $D_{0}$, $J_{1}$, and $D_{1}$ are chosen as 
$J_{1}=-1$, $J_{0}=D_{0}=0$, 
and $D_{1}=-0.1$, $-0.2$, or $-0.3$. 
}
\label{fig15}
\end{figure}

The other three chiral magnets also have 
the same three properties of the specific heat 
as those of the AIAO chiral magnet. 
The results 
for the 3I1O chiral magnet are shown in Figs. \ref{fig15} and \ref{figAdd}(b), 
while the results for the other chiral magnets are not shown.
Then, 
we can understand the origin of each property 
in the same way as for the AIAO chiral magnet 
because 
the optical branch near the $\Gamma$ point 
and the degeneracy of the magnon energies 
do not qualitatively change the specific heat.

\section{Discussion}

We first argue the applicability of the LSWA. 
In the LSWA, 
we express the low-energy effective Hamiltonian 
of a magnetically ordered insulator 
in terms of the quadratic terms of the magnon operators, 
as explained in Sect. 2.3. 
For the arguments about the applicability, 
we need to discuss the effects of the terms neglected in the LSWA. 
The neglected terms can be classified into three parts. 
One part consists of the zeroth-order terms of the magnon operators. 
Since the zeroth-order terms do not usually change the most stable state 
of a three-dimensional magnet, 
their effects will be quantitative. 
Another part consists of the fourth-order and higher-order terms. 
The order of the fourth-order terms is higher than 
that of the terms considered in the LSWA 
by a factor of $\frac{\hat{b}_{ml}^{\dagger}\hat{b}_{ml}}{2S}$. 
Since the factor is small for large $S$ or a low temperature or both, 
the effects of the fourth-order and higher-order terms are negligible 
at low temperatures even for $S=\frac{1}{2}$; 
a low temperature is necessary because at low temperatures 
the number of excited magnons is small. 
Actually, 
this has been demonstrated for an $S=\frac{1}{2}$ collinear antiferromagnet on 
a square lattice~\cite{square-SW}. 
Since the effects tend to be smaller in a three-dimensional system 
than in a two-dimensional system, 
the fourth-order and higher-order terms 
also will not change our main results qualitatively. 
The other part consists of 
the third-order terms. 
The third-order terms are characteristic of noncollinear magnets~\cite{RMP-MagDamp}, 
such as a coplanar antiferromagnet and chiral magnets. 
In contrast to the fourth-order and higher-order terms, 
the third-order terms can affect the magnon properties even at low temperatures. 
However, 
we believe that our main results remain qualitatively unchanged 
even if the third-order terms are taken into account. 
This can be understood as follows. 
First of all, 
the third-order terms, such as 
$\hat{b}^{\dagger}_{ml}\hat{b}^{\dagger}_{nl^{\prime}}\hat{b}_{nl^{\prime}}$, 
connect the ground state to the lowest excited state. 
This suggests that 
the effects of the third-order terms are small for a large energy difference between these states 
and large for a small energy difference. 
In addition, 
the energy difference becomes large or small in the absence or presence of frustration, 
respectively. 
Since the frustration of the Heisenberg interactions is removed 
in the chiral magnets considered in this paper, 
the effects of the third-order terms may be small. 
Actually, 
the energy difference for the AIAO chiral magnet is 
$\Delta E=E_{\textrm{ex}}-E_{\textrm{gs}}=2(J_{1}+4D_{1})-2(-J_{0}-4D_{0})=2J_{0}+8D_{0}$, 
which is large with our parameters 
(e.g., $\Delta E=3.6$ for $J_{0}=1$, $D_{0}=0.2$ and $J_{1}=D_{1}=0$); 
$E_{\textrm{ex}}$ and $E_{\textrm{gs}}$
are the energies of the lowest excited and ground states, respectively. 
Here the lowest excited state for the AIAO chiral magnet is obtained 
by replacing the AIAO spin structures of two of $N$ tetrahedra 
by the 3I1O ones 
because the lowest-energy excitation corresponds to 
the spin flip at the center in Fig. \ref{fig2}(a). 
Similarly, 
we obtain 
the large energy difference for the 3I1O chiral magnet, 
$\Delta E=E_{\textrm{ex}}-E_{\textrm{gs}}=2(-J_{0}-4D_{0})-2(J_{1}+4D_{1})=2|J_{1}|+8|D_{1}|$
(e.g., $\Delta E=3.6$ for $J_{1}=-1$, $D_{1}=-0.2$, and $J_{0}=D_{0}=0$); 
the lowest excited state is obtained by replacing 
the 3I1O/1I3O spin structures 
of two of $N$ tetrahedra 
by the AIAO ones; 
this replacement corresponds to the spin flip at the center in Fig. \ref{fig2}(b). 

We next compare our results with those of several experiments. 
The results for the AIAO or dAIAO chiral magnet 
are compared 
with experimental results for the AIAO type chiral magnet, 
and the results for the 3I1O or d3I1O chiral magnet 
are compared 
with experimental results for the 3I1O type chiral magnet. 
This is because 
it is difficult to distinguish  
between the AIAO and dAIAO chiral magnets 
and between the 3I1O and d3I1O chiral magnets 
in terms of the magnon dispersion and specific heat, 
as shown in Sect. 3.2. 
Our results for the AIAO or dAIAO chiral magnet 
are consistent with 
the experimental results for the AIAO type chiral magnet 
in Sm$_{2}$Ir$_{2}$O$_{7}$ and Nd$_{2}$Ir$_{2}$O$_{7}$: 
the gap at the $\Gamma$ point is consistent with 
the absence of the Goldstone type gapless excitation 
in a neutron scattering measurement on Sm$_{2}$Ir$_{2}$O$_{7}$~\cite{AIAO-exp}; 
the peak of $C_{v}/T$ is consistent with the peak observed experimentally 
for Nd$_{2}$Ir$_{2}$O$_{7}$~\cite{Cv-AIAO}. 
We believe this comparison is meaningful, 
although 
Sm$_{2}$Ir$_{2}$O$_{7}$ and Nd$_{2}$Ir$_{2}$O$_{7}$ 
may have the strong SOC and 
the low-energy effective model may include 
the exchange interactions neglected in our model. 
This is because 
both models include the relevant exchange interactions 
for the AIAO type chiral magnet, 
and because the effects of the neglected interactions 
may be some quantitative modifications. 
Then, 
owing to a lack of experiments on the magnon dispersion 
in the 3I1O type chiral magnet, 
we cannot compare the results for the 3I1O 
or d3I1O chiral magnets 
with the experimental results. 
In addition, 
our results for the specific heat are not comparable 
to the experimental results in
the 3I1O type chiral magnet 
for Dy$_{2}$Ti$_{2}$O$_{7}$~\cite{Cv-3I1O-1,Cv-3I1O-2}. 
This is 
because 
an external magnetic field is absent in our case 
but present in the experiments~\cite{Cv-3I1O-1,Cv-3I1O-2}, 
and because the Zeeman splitting due to the external magnetic field 
causes a Schottky type peak of the specific heat 
as a function of temperature. 

For future experiments, 
we propose that 
if we measure the magnon dispersion of 
the 3I1O type chiral magnet, 
we will observe the gap at the $\Gamma$ point 
and the optical branch near the $\Gamma$ point. 
These observations may hold 
even in the presence of an external magnetic field 
as long as the 3I1O type chiral order is the most stable. 
Thus, by comparing the experimental result with our result, 
we can identify the 3I1O type chiral order.

\section{Summary}

We have studied the magnon dispersion and specific heat 
in four chiral magnets with $\boldQ=\boldzero$ 
at low temperatures using the LSWA for the effective model of the 
$S=\frac{1}{2}$ pyrochlore oxides. 
We have shown that 
the optical branch of the magnon dispersion near $\boldQ=\boldzero$ 
emerges only in the 3I1O and d3I1O chiral magnets, 
while 
the branches are all quasiacoustic 
in the AIAO and dAIAO chiral magnets. 
This is an experimentally distinguishable difference 
between 3I1O type and AIAO type chiral magnets 
and useful for experimentally identifying these chiral orders. 
Also, 
we have shown the gap of the magnon dispersion at $\boldq=\boldzero$ 
in all the chiral magnets, indicating 
the absence of the Goldstone type gapless excitation. 
This is distinct from the property for 
a nonchiral magnet and a characteristic of chiral magnets. 
Then, we have shown that 
the specific heat has no qualitative difference among the four magnets. 
It is thus difficult to determine the kinds of chiral orders from the specific heat.

\begin{acknowledgments}
We thank Prof. S. Maekawa and Dr. J. Ieda for useful discussions. 
The numerical calculations were carried out
using the facilities of the Supercomputer Center, 
the Institute for Solid State Physics, 
the University of Tokyo. 
\end{acknowledgments}

\appendix
\section{Derivation of Eq. (\ref{eq:HLSW})}

We can derive Eq. (\ref{eq:HLSW}) using the Holstein-Primakoff transformation. 
In the Holstein-Primakoff transformation, 
the spin operators 
are expressed in terms of creation and annihilation operators of a magnon as follows:
\begin{align}
&\hat{S}^{\prime z}_{ml}
=S-\hat{b}_{ml}^{\dagger}\hat{b}_{ml},\label{eq:HP-z_LSW}\\
&\hat{S}^{\prime +}_{ml}
=\sqrt{2S-\hat{b}_{ml}^{\dagger}\hat{b}_{ml}}\hat{b}_{ml},\label{eq:HP-+}\\
&\hat{S}^{\prime -}_{ml}
=\hat{b}_{ml}^{\dagger}\sqrt{2S-\hat{b}_{ml}^{\dagger}\hat{b}_{ml}}.\label{eq:HP--}
\end{align}
Equations (\ref{eq:HP-+}) and (\ref{eq:HP--}) can be expanded as follows 
in terms of series of $\hat{b}_{ml}^{\dagger}\hat{b}_{ml}/2S$:
\begin{align}
\hat{S}^{\prime +}_{ml}
=&
\sqrt{2S}
\Bigl(1-\frac{\hat{b}_{ml}^{\dagger}\hat{b}_{ml}}{2S}\Bigr)^{\frac{1}{2}}
\hat{b}_{ml},\label{eq:HP-+_exp}\\
\hat{S}^{\prime -}_{ml}
=&
\hat{b}_{ml}^{\dagger}
\sqrt{2S}
\Bigl(
1-\frac{\hat{b}_{ml}^{\dagger}\hat{b}_{ml}}{2S}
\Bigr)^{\frac{1}{2}}.\label{eq:HP--_exp}
\end{align}
In the LSWA, 
we can approximate Eqs. (\ref{eq:HP-+_exp}) and (\ref{eq:HP--_exp}) 
as the first-order terms of the magnon operator 
because the LSWA includes terms 
up to the second order in the Hamiltonian. 
Namely, 
Eqs. (\ref{eq:HP-+_exp}) and (\ref{eq:HP--_exp}) are approximated as 
$\hat{S}^{\prime +}_{ml}=\sqrt{2S}\hat{b}_{ml}$ and 
$\hat{S}^{\prime -}_{ml}=\hat{b}_{ml}^{\dagger}\sqrt{2S}$.
Thus, $\hat{S}^{\prime x}_{ml}$ and $\hat{S}^{\prime y}_{ml}$ 
are given by
\begin{align}
&\hat{S}^{\prime x}_{ml}
=\sqrt{\frac{S}{2}}(\hat{b}_{ml}+\hat{b}_{ml}^{\dagger}),\label{eq:HP-x_LSW}\\
&\hat{S}^{\prime y}_{ml}
=-i\sqrt{\frac{S}{2}}(\hat{b}_{ml}-\hat{b}_{ml}^{\dagger}).\label{eq:HP-y_LSW}
\end{align}
By using Eqs. (\ref{eq:HP-z_LSW}), (\ref{eq:HP-x_LSW}), and (\ref{eq:HP-y_LSW}), 
we can express Eq. (\ref{eq:Heff-rewrote}) in terms of the magnon operators. 
Then, using their Fourier coefficients 
(e.g., $\hat{b}_{\boldq l}=
\frac{1}{\sqrt{N}}\sum\limits_{m}e^{i\boldq\cdot (\boldR_{m}+\boldr_{l})}\hat{b}_{ml}$), 
we can express Eq. (\ref{eq:Heff-rewrote}) 
as the sum of the zero-order term and second-order terms of the magnon operators 
in the LSWA; 
the zero-order term is given by 
$NS(S+1)\sum_{l,l^{\prime}}M_{ll^{\prime}}^{\prime zz}(\boldzero)$ 
and the second-order terms are given by Eq. (\ref{eq:HLSW}).

\end{document}